\documentclass[%
 preprint,
 amsmath,amssymb,
 aps,
 prb,
 floatfix,
]{revtex4-2}

\usepackage{microtype}
\usepackage{physics}
\usepackage{booktabs}
\usepackage{graphicx}
\usepackage{bm}
\usepackage{xr-hyper}
\usepackage[hidelinks,hypertexnames=false]{hyperref}
\usepackage{bookmark}
\input{si-xr}
\usepackage{tensor}
\usepackage{xcolor}
\usepackage{threeparttable}
\usepackage{chemformula}
\usepackage{float}
\usepackage{placeins}
\usepackage{etoolbox}
\makeatletter
\renewcommand\section{\@startsection{section}{1}{\z@}%
  {0.9ex}%
  {0.55ex}%
  {\normalfont\large\bfseries}}
\renewcommand\subsection{\@startsection{subsection}{2}{\z@}%
  {0.95ex}%
  {0.45ex}%
  {\normalfont\normalsize\bfseries}}
\makeatother
\makeatletter
\AtBeginDocument{%
  \long\def\@makecaption#1#2{%
    \vskip\abovecaptionskip
    \begingroup
      \footnotesize
      \linespread{0.9}\selectfont
      \leftskip\z@ \rightskip\z@ \parfillskip\z@ plus 1fil\relax
      \textbf{#1}\ #2\par
    \endgroup
    \vskip\belowcaptionskip
  }%
}
\makeatother

\makeatletter
\renewcommand{\p@subsection}{}
\newcommand{\BackmatterHeading}[1]{%
  \par\addvspace{2ex \@plus .4ex \@minus .25ex}%
  \noindent\textbf{#1}\par\nobreak
  \addvspace{0.75ex \@plus .2ex \@minus .1ex}%
}
\AtBeginDocument{%
  \renewcommand{\bibsection}{%
    \par\addvspace{2ex \@plus .4ex \@minus .25ex}%
    \noindent\textbf{References}\par\nobreak
    \addvspace{0.75ex \@plus .2ex \@minus .1ex}%
  }%
}
\makeatother

\begin{document}

\citestyle{nature}

\sloppy

\title{Autonomous heterogeneous catalyst discovery with a self-evolving multi-agent digital twin}
\author{Zhilong Song$^{1,2}$}
\author{Zongmin Zhang$^{3}$}
\author{Lixue Cheng$^{1,2,4}$}
\email{lixuecheng@ust.hk}
\affiliation{$^{1}$Department of Chemistry, Hong Kong University of Science and Technology, Kowloon, Hong Kong 999077, China}
\affiliation{$^{2}$IAS Center for AI for Scientific Discoveries, Hong Kong University of Science and Technology, Kowloon, Hong Kong 999077, China}
\affiliation{$^{3}$Department of Computer Science and Engineering, Hong Kong University of Science and Technology, Kowloon, Hong Kong 999077, China}
\affiliation{$^{4}$Department of Chemical and Biological Engineering, Hong Kong University of Science and Technology, Kowloon, Hong Kong 999077, China}

\date{\today}

\begin{abstract}
Theoretical heterogeneous catalysis promises rapid catalyst discovery, yet computational and machine-learning predictions often deviate from experiment and stay confined to narrow material families, for want of a faithful, condition-aware catalytic simulator. We present CatDT (Catalysis Digital Twin), a self-evolving multi-agent system that builds an autonomous digital twin of a working catalyst, unifying gas--solid and liquid--solid modeling. From only a bulk crystal and a natural-language reaction description, eight specialized agents and 27 scientific tools predict stable facets, reconstruct working surfaces, enumerate and rank reaction pathways, locate transition states, and compute kinetics in 5--30~min on a single GPU. Two innovations address the hardest steps: UniMech finds dominant pathways for novel materials at over $10^3\times$ lower cost than exhaustive enumeration by fusing agent-guided proposals with energy-cached graph search, and a memory-augmented reinforcement loop raises barrier-calculation success from 41\% to 84\% across 600 catalytic surfaces. Across seven gas--solid benchmarks---stepped metals, single-atom catalysts, ordered intermetallics, vacancy-rich 2D sulfides and carbides, and a strong-metal--support-interaction (SMSI) interface---every CatDT prediction lies within 0.5--2 times experiment over four orders of magnitude. For propane dehydrogenation, CatDT independently discovers non-precious candidates rivaling the Pt-based industrial benchmark, with a proposed Ni@ZrO$_2$ SMSI overlayer reaching a simulated TOF of $1.63~\text{s}^{-1}$ at $\sim$100\% selectivity. More broadly, the decisive factor for a faithful catalyst digital twin---or any multi-stage scientific simulator---is not raw LLM capability but the engineered harness around it: deterministic tools, persistent memory, and verified self-improvement that compound across models, tools, and runs.
\end{abstract}

\maketitle

\clearpage

\section{Introduction}

Heterogeneous catalysts underpin a large fraction of industrial chemistry\cite{norskov2009towards,norskov2011density}, yet new ones still emerge from the laboratory through an empirical loop of synthesis, characterization, and kinetic testing that has changed little in decades, with months consumed per candidate\cite{zhong2020mlcatalysis}.
An experimentalist holding a new material thus faces a costly choice: iterate this trial-and-error loop, or hand the material to theoretical computation in the hope of faster guidance\cite{norskov2009towards,ulissi2017ncomms}.
The standard recipe that computational catalysis offers in response is itself largely unchanged over the same period.
A low-Miller-index facet is chosen, a static slab is built, candidate adsorbates are placed, and a hand-picked free-energy diagram is derived\cite{norskov2011density,norskov2009towards}.
Activity is then read off a volcano plot\cite{norskov2009towards} or refined through NEB\cite{henkelman2000neb,henkelman2000climbing}, slow-growth MD\cite{cheng2017che}, or Br{\o}nsted--Evans--Polanyi (BEP) and scaling-relation surrogates\cite{logadottir2001bep,bligaard2004bep,abild2007scaling,calle2015scaling_review}.
Each step is a deliberate simplification, but together they leave a systematic gap between the catalyst on paper and the catalyst in the reactor.

Over the past five years, machine learning (ML) has supplied a dedicated replacement for nearly every link of this standard recipe\cite{zhong2020mlcatalysis}.
Underpinning the whole chain, graph neural networks such as SchNet\cite{schutt2018schnet}, DimeNet++\cite{gasteiger2020dimenet}, GemNet\cite{gasteiger2021gemnet}, MACE\cite{batatia2022mace}, and EquiformerV2\cite{liao2023equiformerv2}, trained on the Open Catalyst datasets\cite{chanussot2021oc20,tran2022oc22,shuaibi2025oc25}, reproduce DFT adsorption energies within $\sim$0.1~eV at $\sim$6{,}000$\times$ lower cost, and universal foundation potentials such as UMA\cite{uma2025} extend this fidelity across the periodic table.
On top of this potential, ML has replaced each link of the chain in the order in which the catalyst itself develops.
The first question a real catalyst forces is which crystal facets are exposed, and SurFF\cite{yin2025surff} replaces the arbitrary low-Miller-index pick by ranking all surface free energies and reconstructing the equilibrium Wulff shape.
Under the operating temperature, pressure, electrode potential, and pH the bare surface itself does not remain as cleaved, and VSSR-MC\cite{du2023vssrmc,du2025evssrmc} lifts the static-slab approximation by letting atoms exchange with the gas or electrolyte reservoir until a grand-canonically reconstructed surface emerges.
On this reconstructed surface, the gas-phase species must find their binding sites, and AdsorbML\cite{lan2023adsorbml} and AdsorbDiff\cite{kolluru2024adsorbdiff} replace high-symmetry hand-placement by sampling diverse adsorbate configurations under a universal ML potential.
Bound adsorbates in turn drive a further adsorbate-induced reconstruction, so that downstream energetics reflect the coupled surface--adsorbate system rather than either in isolation.
On the resulting working surface, the reaction unfolds across a network of competing elementary steps, and graph-based reaction-network generators such as CARE\cite{morandi2026care}, Chemoton\cite{unsleber2022chemoton}, and deep-exploration frameworks\cite{zhao2022deepreaction} automate this enumeration in place of a hand-picked free-energy diagram.
Each elementary step crosses a transition state, located $1{,}500\times$ faster by CatTSunami\cite{wander2024cattsunami} for thermal reactions, while CP-MACE\cite{wang2025cpmace} delivers transition-state energetics at constant potential for electrified interfaces.
Finally, the elementary barriers and intermediate free energies are propagated to macroscopic observables by microkinetic solvers such as CatMAP\cite{medford2015catmap}, which accommodates coverage-dependent adsorption energies and lateral adsorbate--adsorbate interactions, or by potential-dependent kinetic Monte Carlo\cite{stamatakis2022kmc}.
Each link is therefore individually machine-learnable.

However, two structural barriers still prevent the assembled ML stack from closing the experiment--theory gap.
The first is \emph{coherence}.
Temperature, pressure, coverage, applied potential, and local pH propagate through every link simultaneously, so facet weighting, surface reconstruction, adsorbate ensemble, transition state, and microkinetic closure are physically meaningful only when referred to the same operating state.
The point-solution tools above each resolve one effect in isolation, and without a system that propagates condition state across them the surface predicted by one tool ceases to be the surface ingested by the next\cite{norskov2011density,wellendorff2012density}.
The second is \emph{scalability}.
Real reaction networks span tens to thousands of intermediates\cite{ulissi2017ncomms,wen2023crn_review}, while existing CRN generators scale combinatorially with chemical space and operate decoupled from the working surface, coverage, and potential under which the network is actually traversed\cite{zhao2021crn_complexity,mou2023bridging}.
On the novel material classes that catalyst discovery most urgently targets, the dominant pathway, taken for granted on textbook transition-metal surfaces, must be re-discovered for each material.
Even once a candidate pathway is identified, the barrier of every elementary step must be located either by nudged elastic band (NEB)\cite{henkelman2000neb,henkelman2000climbing} or by slow-growth molecular dynamics\cite{cheng2017che}, each of which requires a hand-built endpoint geometry consistent with the reconstructed surface and the local coverage.
No general procedure exists for constructing these endpoints automatically, so barrier calculation itself becomes the next combinatorial bottleneck and is currently the most labor-intensive step of any heterogeneous-catalysis study.
What is needed is a digital-twin system\cite{natcompscidt2025} that integrates these heterogeneous tools into a coherent, self-improving pipeline, one that builds realistic surfaces under operating conditions, autonomously discovers and ranks competing pathways, and delivers validated kinetic models with no manual intervention at any link.

Such infrastructure has begun to take shape in adjacent fields.
In chemistry, autonomous LLM agents have rapidly moved from demonstration to deployment: CoScientist plans, executes, and analyzes multi-step laboratory experiments end-to-end\cite{boiko2023coscientist}, and ChemCrow augments a general-purpose LLM with a curated suite of expert chemistry tools to perform organic synthesis, retrosynthesis, and molecular-property prediction\cite{bran2024chemcrow}. Catalyst-Agent extends the same paradigm to high-throughput heterogeneous catalyst screening\cite{catalystagent2025}, while multi-agent frameworks such as CAMEL\cite{li2023camel} further support role-playing collaboration between specialized agents\cite{ramos2025llm_chemistry_review}.
These advances provide a natural substrate for the orchestration layer that heterogeneous catalysis lacks.
At the same time, their primary focus to date has been molecular workflows in solution, and the periodic-slab construction, NEB convergence diagnostics, and multi-scale kinetic coupling that define heterogeneous-catalysis modeling have so far received less attention.
Beyond the choice of scientific scope, whether any such agent achieves scientific reliability rests on the engineering surface around it rather than on the prompts or contexts handed to the model: the tool dispatch, state persistence, deterministic verification, memory, and continuous self-improvement that together constitute the agent \emph{harness}\cite{externalization2026harness,coding_agent_harness2026,nl_agent_harness2026}.
Recent practice has identified the dominant failure modes of long-running agents (context drift, schema misalignment, premature termination) as harness-level rather than model-level pathologies\cite{externalization2026harness,autoharness2026}.
Two harness-level deficiencies in particular remain unaddressed across the chemistry-agent literature: frontier LLMs are surpassed within months, so any agent tightly coupled to a single model ages with it, and current agents are largely stateless, accumulating no operational experience from one run to the next.
The Memento framework\cite{memento2025} points to a way forward by externalizing case-based reasoning, delivering substantial gains without touching the underlying model and thereby providing an inherently model-agnostic, harness-resident self-improvement mechanism.

\begin{figure}[H]
  \centering
  \makebox[\textwidth][c]{\includegraphics[width=1.05\textwidth]{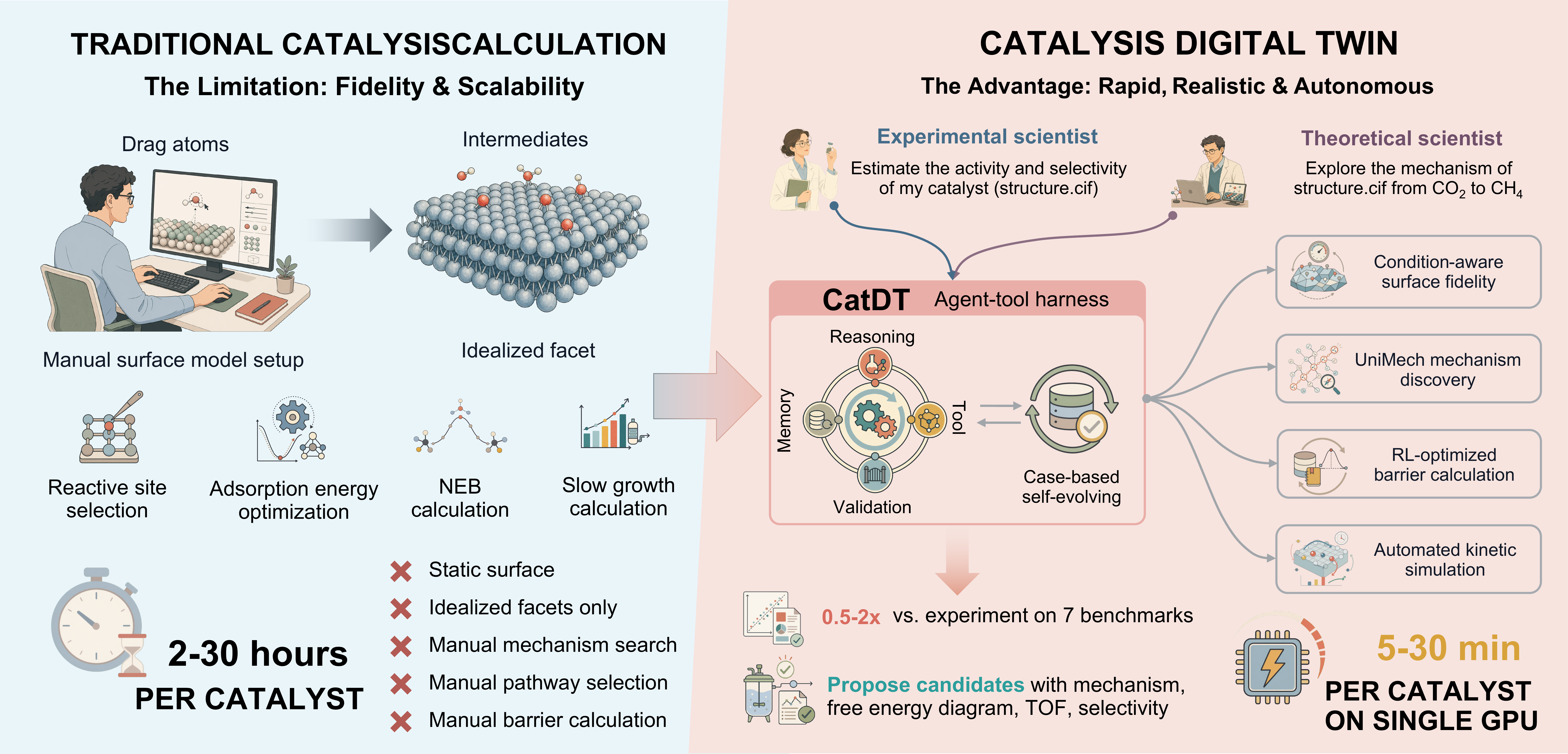}}
  \caption{\textbf{Paradigm shift from traditional catalysis calculation to the CatDT digital twin.}
  \textbf{Left},~Traditional catalysis calculation is limited in fidelity and scalability: the theorist manually builds an idealized facet, places intermediates, and steps through reactive-site selection, adsorption-energy optimization, NEB, and slow-growth calculations, leaving the surface static, the mechanism search manual, and each barrier expert-driven; the full chain requires 2--30~h per catalyst.
  \textbf{Right},~CatDT delivers a rapid, realistic, and autonomous digital twin for both experimental and theoretical scientists. An agent--tool harness with memory-based self-evolution cycles reasoning, tool execution, validation, and cross-run memory to provide condition-aware surface fidelity, UniMech mechanism discovery, reinforcement-learning-optimized barrier calculation, and automated kinetic simulation, returning candidates with mechanism, free-energy diagram, TOF, and selectivity in 5--30~min per catalyst on a single GPU and matching experiment within 0.5--2$\times$ on seven benchmarks.}
  \label{fig:overview}
\end{figure}

Here we introduce CatDT (Catalysis Digital Twin), a self-evolving multi-agent system in which deterministic tools, cross-run memory, and verification gates wrap the agent layer (Fig.~\ref{fig:overview}).
This harness lets an LLM drive condition-dependent surface modeling, automated mechanism discovery, and self-improving transition-state construction in a single autonomous pipeline.
Eight specialized agents coordinating 27 scientific tools, built on CAMEL\cite{li2023camel} under strict agent--tool separation, carry the chain from a bulk crystal and a natural-language reaction description to microkinetic observables with no manual intervention.
Two innovations target the key challenges: UniMech (Agent~M1) constructs reaction networks and discovers energy-ranked pathways at over $10^3\times$ lower cost than exhaustive enumeration\cite{morandi2026care}, and a memory-augmented reinforcement loop lifts transition-state endpoint success from 41\% to 84\% across 600 diverse catalytic surfaces\cite{chanussot2021oc20,memento2025}.
At every step of the pipeline, the agents reason about each result rather than calling tools and accepting their outputs: when an NEB run fails to converge or an endpoint is rejected by the deterministic gate, the validation auditor diagnoses overlap, element-count, or path-collision causes and proposes targeted geometric corrections that feed the next iteration.
On seven gas--solid benchmarks (stepped metals, single-atom catalysts, ordered intermetallics, vacancy-rich two-dimensional sulfides, two-dimensional carbides, and a strong-metal--support-interaction, SMSI, interface), every CatDT prediction lies between 0.5 and 2 times the corresponding experimental value across measurements spanning four orders of magnitude.
The same reflection drives CatDT to autonomously propose a class of non-precious propane-dehydrogenation catalysts, including cross-family designs absent from prior literature, that rival the Pt-based industrial benchmark. The agent-proposed Ni@ZrO$_2$ SMSI overlayer reaches a CatMAP--KMC propylene TOF of $1.63~\text{s}^{-1}$ at $\sim$100\% selectivity, about five times the PtSn industrial reference.
Because the agent layer is decoupled from the tool layer, CatDT grows with each advance in foundation models and scientific tools rather than ageing alongside any one of them.


\section{Results}

\subsection{CatDT Architecture}

Running the chain introduced above end-to-end currently takes a researcher 6--30~h of expert labor per catalyst and is hard to reproduce or scale.
CatDT replaces that serial, human-intensive workflow with a closed-loop multi-agent system that completes the same pipeline in 5--30~min on a single GPU, with full traceability and no manual intervention (Fig.~\ref{fig:overview}).
Eight specialized agents are orchestrated within a CAMEL\cite{li2023camel}-based system around one principle: agents reason, tools compute. Every agent therefore acts only by invoking deterministic scientific tools (Table~\ref{si-tab:tools_full}) and never computes a physical quantity itself.
This strict separation makes predictions reproducible despite agent stochasticity, lets the reasoning backend be replaced without touching any calculation, and lets each tool be validated and upgraded independently.
This is harness engineering for the physical sciences\cite{externalization2026harness,coding_agent_harness2026}: feedforward guides (typed schemas, staging plans, and deterministic preconditions) steer each agent call before it is made, while feedback sensors (programmatic gates, energy screens, NEB-convergence diagnostics, and a persistent cross-run memory) turn each outcome into a durable fix applied once at the harness level rather than re-prompted at every run.
Without this discipline, a stochastic multi-agent system does not yield the condition-aware predictions a real-catalyst digital twin requires.

\begin{figure}[H]
  \centering
  \makebox[\textwidth][c]{\includegraphics[width=1.05\textwidth]{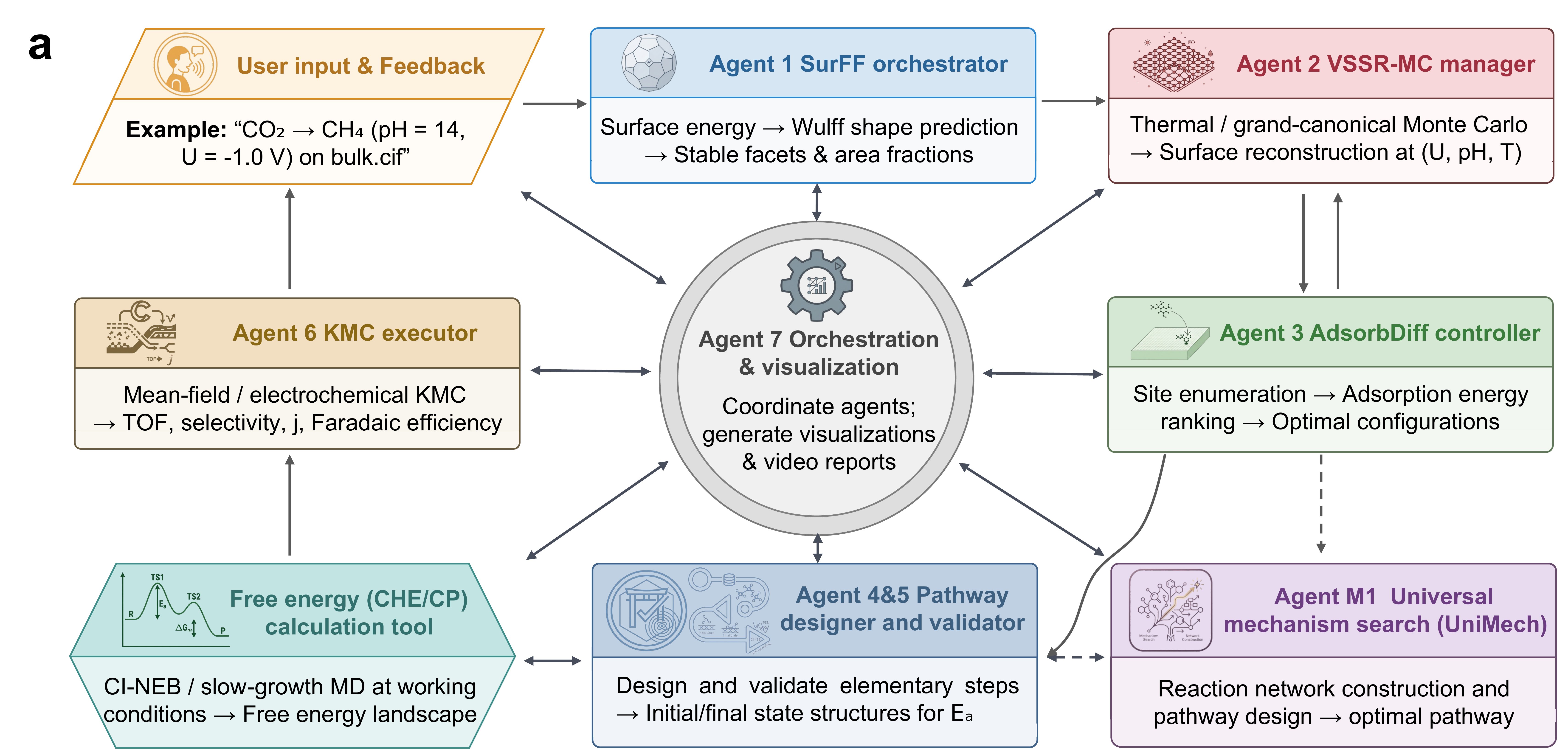}}\\[4pt]
  \makebox[\textwidth][c]{\includegraphics[width=1.05\textwidth]{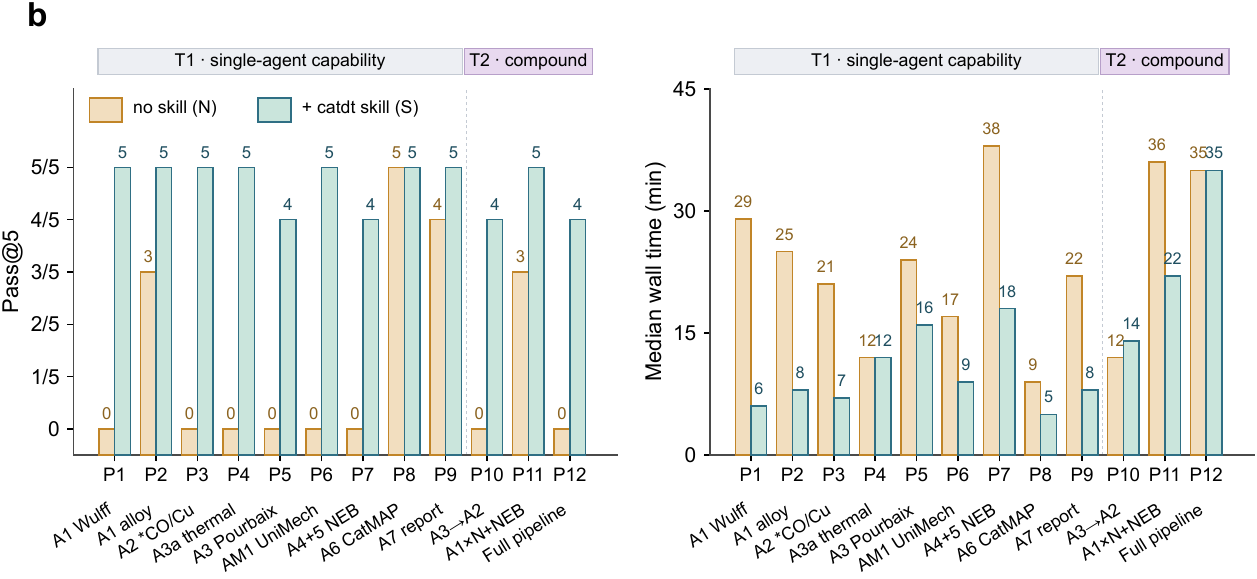}}
  \caption{\textbf{CatDT hub-and-spoke multi-agent architecture and skill ablation.}
  \textbf{a}.~Agent~7 (centre) coordinates eight specialized agents: Agent~1 (SurFF) for Wulff-shape prediction and facet ranking, Agent~2 (VSSR-MC, two-step: clean reconstruction then adsorbate-induced reconstruction) for condition-dependent surface modeling at specified $(T, P, U, \text{pH})$, Agent~3 (AdsorbDiff) for adsorption-site sampling, Agent~M1 (UniMech) for multi-pathway enumeration and energy-guided pruning, Agent~4 (Geometry engine + Memento memory) for elementary-step endpoint design, Agent~5 (deterministic gates + energy screen) for validation, and Agent~6 (CatMAP / electrochemical kMC) for microkinetic modeling.
  A standalone barrier-calculation tool provides CI-NEB or slow-growth MD under working conditions.
  User input: a bulk structure file and a natural-language reaction description.
  \textbf{b}.~Ablation of the CatDT coding-agent skill across 12 prompts spanning single-agent capability (T1) and compound orchestration (T2), 5 runs per arm. N, no skill loaded; S, catdt skill loaded. Left, scientifically-sound pass@5 (a run counts only if it returns the requested deliverable computed with a real ML/DFT potential and physically correct energetics); right, median wall time. Scoring rubric, per-prompt results, and failure taxonomy in Supplementary Section~\ref{si-note:skill_ablation} (Table~\ref{si-tab:skill_ablation}).}
  \label{fig:architecture}
\end{figure}

The multi-agent architecture adapts its tool backends to the catalytic interface under study (Fig.~\ref{fig:architecture}).
For gas--solid (thermal) interfaces, the pipeline begins by building a realistic working surface.
Agent~1 calls SurFF\cite{yin2025surff} to predict surface free energies for all Miller-index slabs with an EquiformerV2\cite{liao2023equiformerv2} backbone, constructs the equilibrium Wulff shape, and returns the top-$N$ exposed facets as simulation-ready slabs.
Agent~2 then runs VSSR-MC\cite{du2023vssrmc} at the operating $(T,P,U,\mathrm{pH})$ in two coupled passes, thermally reconstructing the bare slab and then re-reconstructing it once the adsorbates are placed, so that downstream energetics reflect the coupled surface--adsorbate system.
Between the two passes, Agent~3 samples adsorbate configurations on the reconstructed surface with AdsorbDiff\cite{kolluru2024adsorbdiff}, a conditional denoising diffusion model that captures non-symmetric binding minima missed by high-symmetry enumeration (Supplementary Section~\ref{si-note:architecture}, Tables~\ref{si-tab:tools_full}--\ref{si-tab:agents}).

On this working surface, Agent~M1 (UniMech) takes over the second half of the chain by enumerating competing reaction pathways and emitting an energy-ranked shortlist of elementary steps directly from the natural-language reaction description, without any human-supplied mechanism input. The full design of this module is described in Section~\ref{subsec:unimech}.
For each elementary step on the shortlist, an iterative design--validation loop (Agents~4 and~5) constructs the NEB-ready endpoint geometries that are by far the most computationally demanding component of the pipeline.
Agent~4 (Pathway Designer) takes baseline tool outputs, a pre-computed staging plan that pins hollow- or bridge-site positions for atoms entering or leaving the surface, retrieved cases and distilled knowledge/skill items from a persistent memory bank, and feedback from Agent~5, and outputs a structured specification of the three-dimensional positions of all moving atoms while keeping the slab geometry fixed.
Agent~5 (Validation Auditor) applies a hybrid gate combining deterministic programmatic checks (element-count consistency, atomic overlap ${<}$0.8~\AA, interpolated-path collision detection), a UMA pre-NEB energy screen, and a reasoning-based geometric plausibility review.
Fatal programmatic or energy-gate violations enforce a \textsc{fail} verdict deterministically, whereas reasoning-only objections are retained as advisory feedback for redesign, and only structures that clear the deterministic gate advance to barrier calculation, with the loop iterating up to 10 times (Supplementary Section~\ref{si-note:architecture}; Tables~\ref{si-tab:tools_full}--\ref{si-tab:agents}).
The memory bank, the supervisory bandit policy, and the reinforcement signal that together enable this loop to improve across runs are detailed in Section~\ref{subsec:memory_rl}.

Once every elementary step passes validation, a deterministic barrier tool runs two-phase climbing-image NEB on the UMA\cite{uma2025} universal ML potential (Methods). Any barrier outside a physically plausible window triggers a return to Agent~5 for targeted redesign, closing an outer quality-control loop.
The validated barriers feed microkinetic modeling (Agent~6), which invokes CatMAP\cite{medford2015catmap} to compute turnover frequencies, surface coverages, selectivities, and rate-determining steps, closing the chain from atomic structure to macroscopic catalytic performance.
Throughout, Agent~7 acts as central coordinator: it sequences all agents through a workflow state machine, manages checkpoints, and produces energy diagrams, structure visualizations, and the final report.

For liquid--solid (electrocatalytic) interfaces, the same agent orchestration is retained, but three tool backends are swapped to account for the electrochemical environment (Fig.~\ref{fig:architecture}).
Agent~2 switches to the electrochemical extension of VSSR-MC\cite{du2025evssrmc}, which operates in the grand-canonical ensemble at fixed electrode potential $U$ and solution pH and lets the surface exchange electrons and protons with the reservoir, thereby capturing potential-driven reconstruction (oxide formation, hydroxylation, metal dissolution) that is absent under thermal conditions.
The UMA-based CI-NEB is replaced by slow-growth molecular dynamics with explicit solvent under the computational hydrogen electrode (CHE) framework\cite{cheng2017che,wang2025cpmace}, because electrocatalytic barriers depend on the applied potential, the local electrostatic environment, and solvent reorganization at the electrode--electrolyte interface. Energies are evaluated with a constant-potential machine-learning potential trained for the target electrochemical interface (see Section~\ref{subsec:ml_potentials}; Supplementary Section~\ref{si-note:architecture}; Tables~\ref{si-tab:tools_full}--\ref{si-tab:agents}).
Finally, Agent~6 switches from CatMAP to electrochemical kinetic Monte Carlo (kMC), which propagates potential-dependent rate constants to predict current densities, Faradaic efficiencies, and selectivity--potential maps.
Because the agent layer is decoupled from the tool layer, extending CatDT to other catalytic environments (e.g.\ solid--solid interfaces, photocatalysis) requires only registering additional tool backends without modifying any agent logic.
The next two subsections complete this chain: Agent~M1 (UniMech) replaces user-supplied mechanisms with an energy-ranked pathway shortlist on the working surface, and a memory-augmented Agent~4/5 loop converts each ranked step into a converged NEB barrier without per-step human intervention.

To quantify what this agent--tool harness adds beyond a capable general-purpose coding agent acting alone, we ablated the CatDT skill across twelve catalysis prompts spanning single-agent capability (T1) and compound orchestration (T2), running each prompt five times in two arms---without the skill (N) and with it (S)---and scoring a run as a pass only when it is \emph{scientifically sound}: it returns the requested deliverable computed with a genuine machine-learned or first-principles interatomic potential and physically correct energetics, rather than merely terminating without error (Fig.~\ref{fig:architecture}b; Supplementary Section~\ref{si-note:skill_ablation}, Table~\ref{si-tab:skill_ablation}).
Without the skill a frontier coding agent is sound on only 15 of 60 runs. Among the ten prompts that demand an atomistic potential it succeeds on just two---an alloy surface-energy case and the multi-facet barrier comparison---and only in the reps where it manages to load a community ML potential (MACE) by itself; elsewhere it falls back on empirical EMT, hand-coded lattice-gas Hamiltonians, or hardcoded literature energies, produces no barrier at all on the step that pins the universal potential by name (P7, gated and unreachable), and resorts to surrogate barriers in the full pipeline (P12).
Its most reliable successes are the two tasks that need no atomistic calculation at all---mean-field microkinetics from energetics supplied in the prompt, and HTML aggregation of existing output files.
Loading the skill raises the sound pass rate to 56 of 60 runs---5/5 on eight of the twelve prompts and at least 4/5 on every prompt---because each task is dispatched to the same deterministic, independently validated tools, a universal ML potential, and the orchestration loop described above, at comparable or lower wall time (for example, the three-facet barrier comparison drops from a 36- to a 22-min median).
The complete per-prompt scoring for both arms, the sound-gating rubric, and the failure taxonomy are given in Supplementary Section~\ref{si-note:skill_ablation} (Table~\ref{si-tab:skill_ablation}).

\FloatBarrier

\subsection{Agent~M1 (UniMech): Universal Mechanism Search with Energy-Guided Pruning}\label{subsec:unimech}

The architecture above resolves the coherence bottleneck but still assumes a single, pre-specified mechanism per reaction, an assumption that breaks on two fronts.
Even for canonical reactions (CO$_2$RR, OER, HER, NRR, methanol synthesis), real catalytic systems host branching networks of tens to hundreds of competing routes whose relative weights depend sensitively on facet, coverage, and reconstruction\cite{ulissi2017ncomms,wen2023crn_review}.
For emerging material classes (single-atom alloys, vacancy-engineered chalcogenides, ordered intermetallics, SMSI overlayers), even the dominant pathway is unknown \emph{a priori}: the textbook mechanism may be inactive, a non-canonical intermediate may govern selectivity, or the surface may favor a route that has no analog in the metal-catalysis literature.
Automated chemical reaction-network (CRN) construction has emerged as the key response to this scalability problem\cite{wen2023crn_review}: graph-theoretical and autonomous frameworks have systematized pathway discovery in solution\cite{unsleber2022chemoton,hashemi2022renegat,steiner2022homhet}, and on heterogeneous interfaces deep-exploration approaches\cite{zhao2022deepreaction,zhao2021crn_complexity}, ML-bridged microkinetics\cite{mou2023bridging}, automated mechanism generation\cite{kreitz2023autocrn}, and most recently the end-to-end CARE framework\cite{morandi2026care} have demonstrated that complete networks of up to 40{,}000 intermediates and 455{,}000 reactions can be enumerated and screened against experimental trends.
CARE's strength is exhaustive full-CRN generation: it enumerates competing intermediates and product channels, then accelerates screening with a barrier-prediction network trained on adsorbate configurations on pure-metal surface structures, enabling selectivity analysis once the full network is scored.
Two limitations nevertheless persist across these CRN approaches.
First, exhaustive full-network workflows must score every intermediate and elementary reaction before any specified product pathway can be ranked, so the cost scales combinatorially with chemical-space size even when scoring is accelerated by learned predictors.
Second, template-based generators have a restricted element scope. CARE, for example, covers only C/H/O/N and excludes the sulfur-, phosphorus-, halogen-, and metal-containing adsorbates central to hydrodesulfurization, chlor-alkali, and biomass conversion.

\begin{figure}[H]
  \centering
  \makebox[\textwidth][c]{\hspace*{0.055\textwidth}\includegraphics[width=1.0\textwidth]{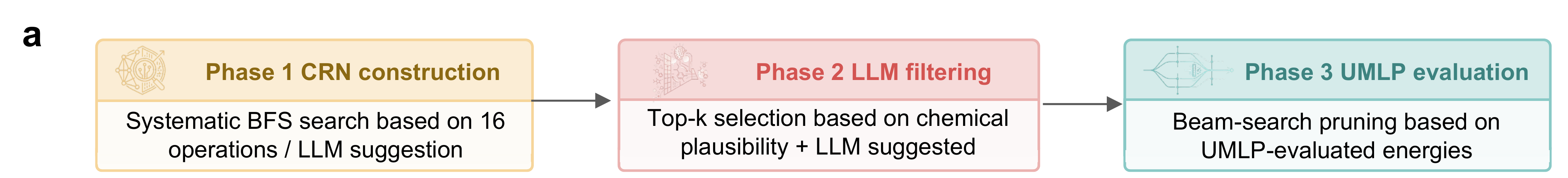}}\\[2pt]
  \makebox[\textwidth][c]{\includegraphics[width=1.1\textwidth]{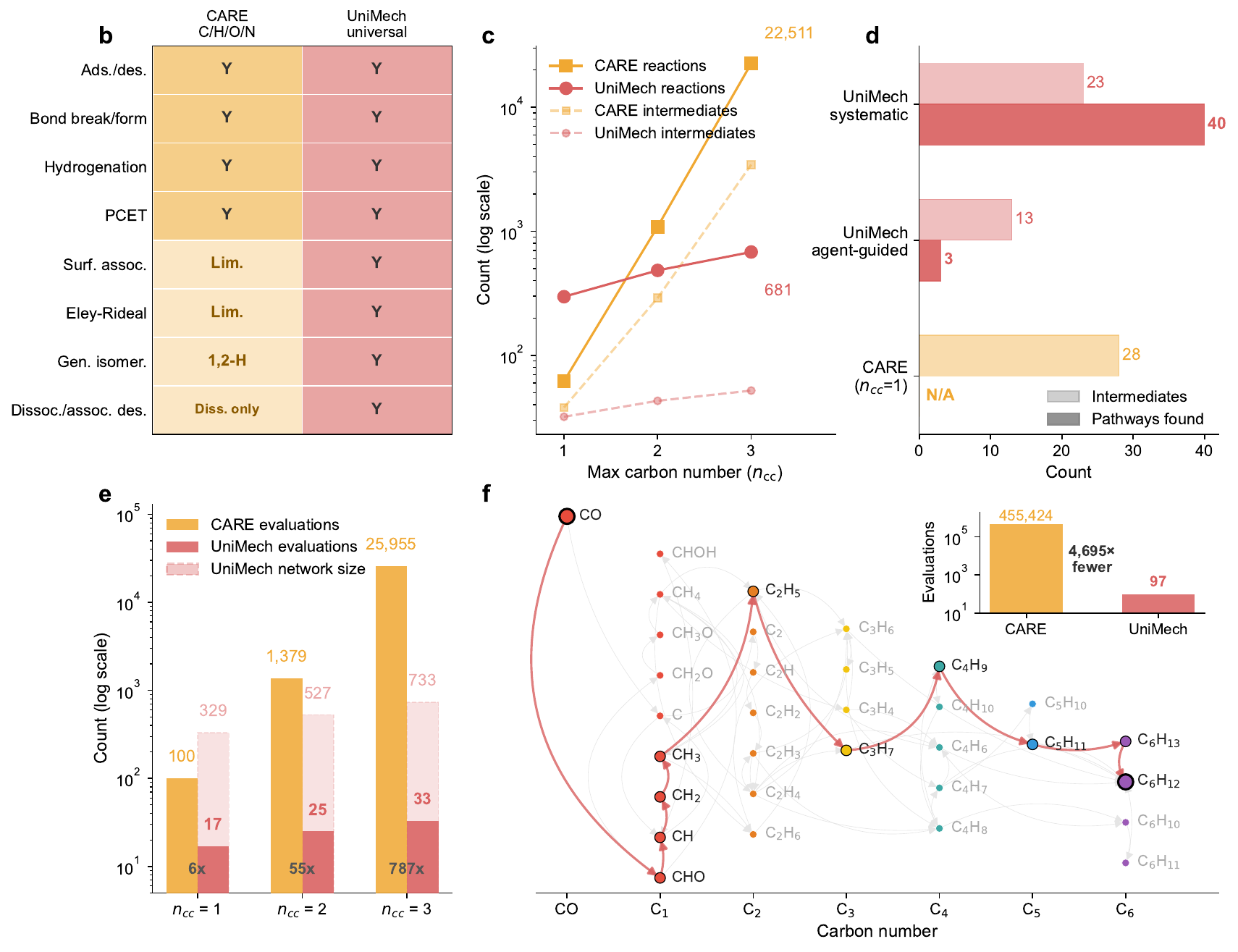}}
  \caption{\textbf{Agent~M1 (UniMech) architecture and benchmark against CARE.}
  \textbf{a}.~Overview of the Agent~M1 mechanism search workflow within the CatDT pipeline (see Fig.~\ref{si-fig:si_unimech_detail} for the expanded view).
  \textbf{b}.~Elementary operation scope with abbreviated panel labels. CARE is restricted to C/H/O/N. Its last four classes are limited rather than general (``Lim.'', restricted template support, ``1,2-H'', only 1,2-H shifts, ``Diss. only'', dissociative adsorption without the general associative-desorption class).
  \textbf{c}.~Network scaling with chemical-space complexity ($n_\text{cc}$): exhaustive enumeration (CARE) vs.\ beam-pruned search (UniMech).
  \textbf{d}.~CO$_2$$\to$CH$_4$ pathway discovery on Cu(211): agent-guided mode identifies major mechanism families; systematic mode discovers additional pathway variants. Bar color encodes engine (red, UniMech; amber, CARE); light tint, intermediates; dark tint, pathways found.
  \textbf{e}.~Energy evaluation cost comparison. Solid bars give the actual evaluation count; the translucent UniMech bar gives its total network size (beam-search frontier $\ll$ network). CARE must score every node and edge, so its network size equals its evaluation count and is shown by the single solid bar.
  \textbf{f}.~Fischer--Tropsch pathway network from *CO to *C$_6$H$_{12}$ on Ni(111). Nodes colored by carbon number; red trace: top-ranked carbide pathway.}
  \label{fig:unimech_benchmark}
\end{figure}

CatDT addresses both limitations through Agent~M1, termed UniMech (Universal Mechanism Search Engine), a goal-directed, element-agnostic pathway-discovery module that replaces exhaustive enumeration with a hybrid generation--search architecture (Fig.~\ref{fig:unimech_benchmark}a).
Two complementary candidate generators run in parallel: an \emph{agent-guided generator} lets the foundation model propose chemically motivated pathways directly from the natural-language reaction specification, drawing on the literature priors encoded in the LLM to recover canonical mechanisms in a single call, while a \emph{systematic generator} expands the reaction graph through sixteen directional element-agnostic bond operations implemented as RDKit manipulations (Methods, Supplementary Section~\ref{si-note:unimech}; Fig.~\ref{si-fig:si_unimech_detail}; Table~\ref{si-tab:unimech_ops}), so the same codebase covers C/H/O/N together with sulfur-, halogen-, phosphorus-, and metal-containing adsorbate chemistries.
Their union feeds a unified tree-search backend (best-first beam search by default, with a switch to Monte Carlo Tree Search under UCB1 selection for non-monotonic free-energy landscapes in which the lowest-barrier path passes through transiently unfavorable intermediates, see Methods) that emits energy-ranked complete pathways.

Both expansion modes share the same energy cache, which evaluates every unique adsorbed species at most once on the reconstructed surface delivered by Agent~2, so the cost of adding an additional pathway scales with the number of \emph{new} intermediates it introduces rather than with its length. Cache entries store Gibbs free energies, with zero-point, vibrational, and ideal-gas-entropy corrections drawn at the operating $(T,P)$ from the CatDT 569-species molecule database constructed for this work (Methods, Supplementary Section~\ref{si-note:thermo_db}, Fig.~\ref{si-fig:thermo_db_stats}, Table~\ref{si-tab:thermo_db_composition}).

UniMech executes this design in three phases.
Phase~1 (candidate generation) runs the agent-guided and systematic generators in parallel. For CO hydrogenation to C$_6$H$_{12}$ on Ni(111), the systematic generator alone discovers ${\geq}$200 distinct pathways in under 30~s with no energy evaluation, while the agent-guided generator simultaneously surfaces literature-known mechanism families.
Phase~2 (chemical plausibility filtering) prunes valence-implausible intermediates with a generic RDKit-based check and lets the LLM rank survivors by chemical reasonableness, retaining the top-10 most promising routes.
Phase~3 (energy-guided tree search) hands these candidates to the cache-shared beam or MCTS backend, which evaluates only the frontier intermediates with UMA, prunes siblings whose adsorption-energy gap exceeds $\Delta_\text{prune} = 0.30$~eV, and emits a ranked shortlist of complete, energy-annotated pathways ready for Agent~4/5 endpoint construction and NEB.

We benchmark UniMech against CARE in five complementary respects that move from chemical expressivity, through scaling behavior, to performance on representative reactions (Fig.~\ref{fig:unimech_benchmark}b--f), while keeping the task distinction explicit.
CARE is an exhaustive full-CRN generator: after a full network is enumerated and the elementary reactions are scored by its pure-metal-surface-trained barrier-prediction network, the resulting microkinetic model can compare competing reactions, product channels, and selectivity within the enumerated C/H/O/N domain.
UniMech instead performs target-directed search, asking for low-energy routes to specified products and spending energy evaluations only on the frontier states needed to rank those routes.
Through element-agnostic RDKit bond manipulations, UniMech covers all eight generalized operation categories shown in Fig.~\ref{fig:unimech_benchmark}b.
CARE's blueprint templates cover a narrower C/H/O/N subset: adsorption/desorption, bond dissociation/formation, hydrogenation/PCET, limited 1,2-H rearrangement, and H$_2$/O$_2$/N$_2$ dissociative adsorption are available, whereas the last four operation classes in Fig.~\ref{fig:unimech_benchmark}b remain restricted rather than general-purpose.

Given comparable coverage, we next compare how each engine scales with chemical-space complexity.
As complexity grows from $n_\text{cc}=1$ to $n_\text{cc}=3$, CARE's exhaustive enumeration grows combinatorially from 62 to 22{,}511 reactions, because every bond type in every molecule is recursively broken (Fig.~\ref{fig:unimech_benchmark}c).
UniMech's energy-guided tree search instead generates only 297--681 candidates across the same range, evaluating frontier nodes at each depth and pruning siblings whose adsorption energies exceed the best candidate by more than $\Delta_\text{prune} = 0.30$~eV.

To verify that this controlled growth does not compromise mechanistic coverage on a canonical reaction, we apply both engines to CO$_2\to$CH$_4$ on Cu(211) (Fig.~\ref{fig:unimech_benchmark}d).
The agent-guided mode identifies three mechanism families in a single API call: the COOH*-mediated route, the carbide pathway through *C, and the formate pathway via *HCOO, collectively spanning the major CO$_2$RR branches known from the literature.
The systematic mode independently uncovers 40 additional pathway variants through recursive bond enumeration, including the HCOOH intermediate that the LLM proposal missed, so the two modes provide complementary, non-redundant coverage.
CARE produces 28 adsorbed intermediates for the same C$_1$O$_2$ chemical space. Pathway ranking and selectivity then come from scoring the full CRN with its downstream barrier-prediction network.

Beyond the count of pathways, the practical cost is the number of energy evaluations needed to rank them (Fig.~\ref{fig:unimech_benchmark}e).
CARE must evaluate every intermediate and every reaction before microkinetic modeling can begin, rising from 100 evaluations at $n_\text{cc}=1$ to 25{,}955 at $n_\text{cc}=3$, whereas UniMech evaluates only 17--33 frontier states across the same range.
This 6--787$\times$ reduction is driven jointly by the shared energy cache, which eliminates redundant evaluations of intermediates appearing on multiple pathways, and by sibling-comparison pruning, which discards thermodynamically unfavorable branches before they propagate downstream.

These advantages converge on the C$_6$O$_1$ Fischer--Tropsch system, CARE's flagship application and the most demanding case in this benchmark (Fig.~\ref{fig:unimech_benchmark}f).
CARE generates 39{,}893 intermediates and 455{,}424 elementary reactions, requiring 4{,}130~s for blueprint construction alone on 16 CPU cores plus an additional 4~h of GAME-Net-UQ barrier prediction on a 24-core CPU with GPU\cite{morandi2026care}.
UniMech, starting from *CO on a 48-atom Ni(111) slab under Fischer--Tropsch conditions (523~K, 10~bar, H$_2$/CO\,=\,2), discovers complete energy-ranked pathways to both C$_6$H$_{12}$ (hexene) and C$_6$H$_{14}$ (hexane) using only 97 UMA evaluations in 635~s, a reduction of more than three orders of magnitude.
The top-ranked C$_6$H$_{12}$ pathway (Fig.~\ref{fig:unimech_benchmark}f, red trace) proceeds via the carbide mechanism in ten elementary steps: CO is hydrogenated to formyl, whose C--O bond then cleaves to a surface methylidyne and is hydrogenated to a methyl monomer (*CO\,$\to$\,*CHO\,$\to$\,*CH\,$\to$\,*CH$_2$\,$\to$\,*CH$_3$); the surface alkyl then grows one carbon at a time through successive C$_1$ (CH$_2$) insertions (*CH$_3$\,$\to$\,*C$_2$H$_5$\,$\to$\,*C$_3$H$_7$\,$\to$\,*C$_4$H$_9$\,$\to$\,*C$_5$H$_{11}$\,$\to$\,*C$_6$H$_{13}$), and terminates by $\beta$-hydride elimination to 1-hexene.
This chain-growth pattern, in which C$_{n+1}$ species form by addition of a C$_1$ fragment to the C$_n$ alkyl, is the canonical Anderson--Schulz--Flory carbide mechanism for Fischer--Tropsch synthesis\cite{vansanten2013ftmech,foppa2019ch2monomer}.
With pathway enumeration thus reduced to a standard pipeline step, the remaining bottleneck is converting each ranked step into a converged NEB transition state on the same slab, the subject of the next subsection.
Supplementary inventories for CO$_2\to$CH$_4$ on Cu(211), hierarchical summaries for Fischer--Tropsch scaling, CARE-vs.-UniMech evaluation-cost breakdowns, and phase-one CARE statistics are collected in Supplementary Section~\ref{si-note:unimech} (Tables~\ref{si-tab:pathways_ch4}--\ref{si-tab:pathways_summary}; Figs.~\ref{si-fig:si_ch4_network}--\ref{si-fig:si_phase1_care}).

\FloatBarrier

\subsection{Self-Improving Endpoint Construction via Memory-Augmented Reinforcement Learning}\label{subsec:memory_rl}

UniMech now delivers an energy-ranked shortlist of elementary steps on the reconstructed working surface, with every step energy already corrected for ZPE, vibrational enthalpy, and entropy at the operating $(T,P)$ from the CatDT molecule database introduced in the previous section.
What remains is to turn each $\Delta G$-ranked step into a converged NEB barrier, which requires geometrically valid initial and final endpoints on the same slab.
This is the most labor-intensive link of the entire pipeline and the one where it most often fails: atomic overlaps, element-count mismatches, and geometrically implausible configurations are common failure modes on the first attempt, and a stateless agent would repeat the same mistakes indefinitely.

Agents~4 and~5 construct NEB-ready endpoints in two coupled geometric stages on the reconstructed slab from Agent~2.
The first stage \emph{builds the product from the reactant}. Starting from the relaxed adsorbed reactant, which is either the previous step's relaxed product for sequential steps or the initial adsorbed surface for the first step, Agent~4 reads the element delta between the current and target intermediates and adds, removes, or rearranges the moving atoms while keeping the slab geometry immutable, using the target molecule's gas-phase coordinates from the CatDT molecule database as the bonding template.
The second stage \emph{adds the staging atoms required by NEB}. Because the nudged elastic band demands identical element multisets on both endpoints, any atom present on only one side is placed on the opposite endpoint at a chemically reasonable hollow or bridge site, drawn from a pre-computed geometry-based staging plan that RMSD-matches reactant and product adsorbate atoms and pins the unmatched ones to the nearest-neighbor sites of their bond partners.
An overlap-corrected interpolation, which starts from a linear path and iteratively pushes apart atoms that are closer than their interpolated target distance, then yields the NEB initial frames, which Agent~5 screens for atomic overlaps below 0.8~\AA, element-count consistency, and interpolated-path collisions before any barrier calculation is attempted.
Full pseudocode for both stages, the path-interpolation algorithm, and the deterministic validation criteria is provided in Supplementary Section~\ref{si-note:rl_loop} (Secs.~\ref{si-ssec:two_stage}--\ref{si-ssec:rl_ablation}; Figs.~\ref{si-fig:si_case_bank}--\ref{si-fig:si_pos_neg_ablation}; Tables~\ref{si-tab:rl_ablation_full}--\ref{si-tab:ablation}).

\begin{figure}[H]
  \centering
  \makebox[\textwidth][l]{\hspace*{-0.025\textwidth}\includegraphics[width=1.02\textwidth]{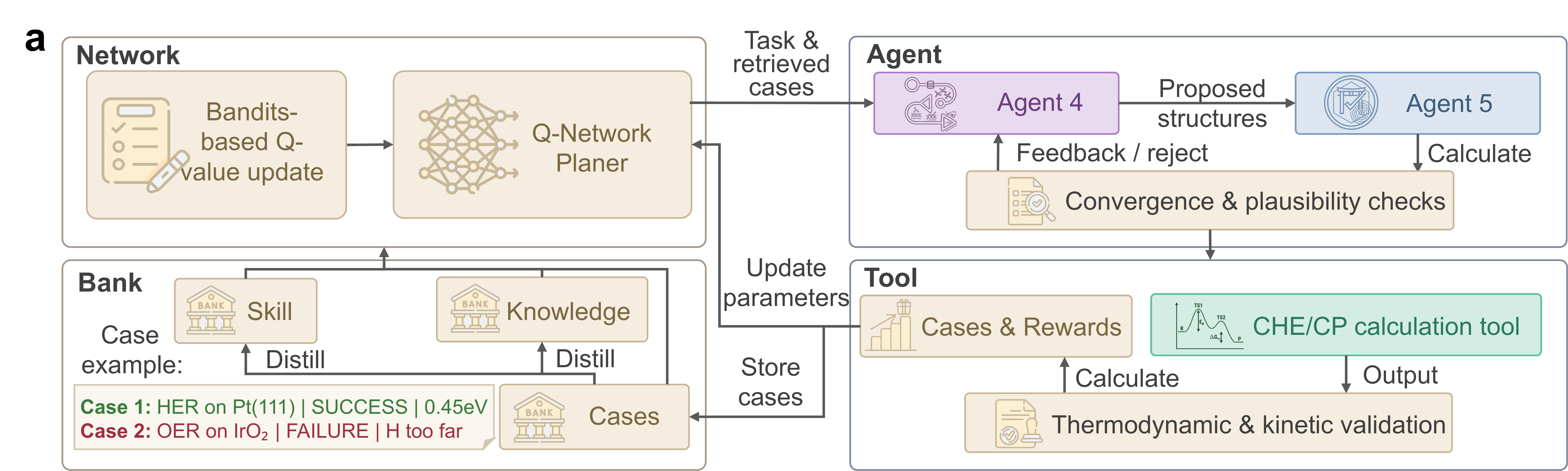}}\\[2pt]
  \makebox[\textwidth][c]{\includegraphics[width=1.05\textwidth]{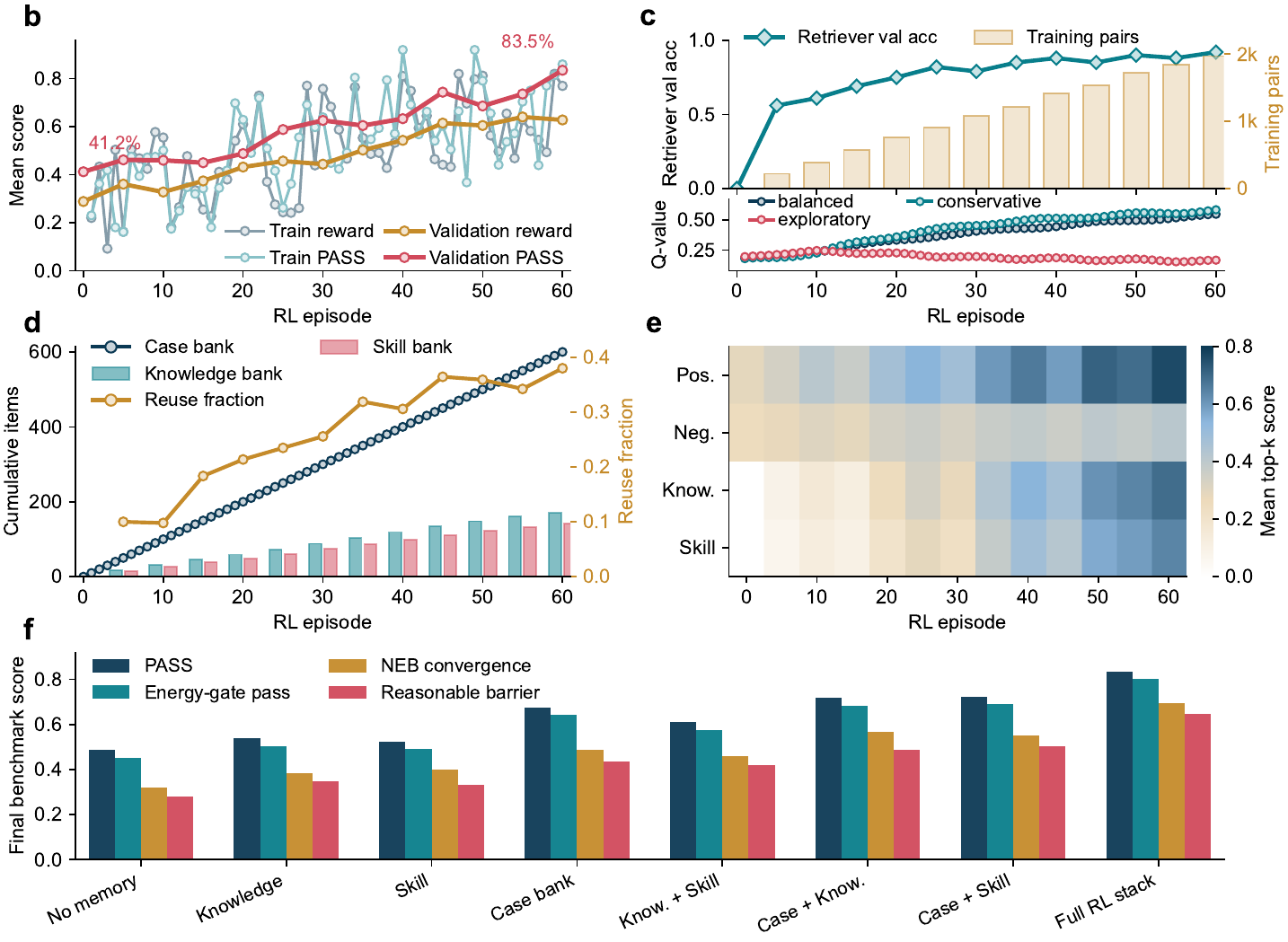}}
  \caption{\textbf{Memory-augmented reinforcement loop for Agents~4 and~5.}
  \textbf{a}.~Architecture, organised into four interacting modules: \emph{Network} (workflow bandit and Q-network planner), \emph{Bank} (cases plus distilled knowledge and skill items), \emph{Agent} (Agent~4 endpoint proposer coupled to Agent~5 deterministic validator), and \emph{Tool} (CHE/CP barrier tool with thermodynamic and kinetic validation).
  \textbf{b}.~Train and held-out validation reward and PASS rate over 60 RL episodes.
  \textbf{c}.~Parametric-retriever validation accuracy with training-pair growth (top); $\varepsilon$-greedy Q-values for the three workflow strategies (bottom).
  \textbf{d}.~Case-bank, KnowledgeBank, and SkillBank size and reused-case fraction over training.
  \textbf{e}.~Mean top-$k$ retrieval-score heatmap across validation checkpoints for positive cases, negative cases, knowledge items, and skill templates.
  \textbf{f}.~Staged ablation of PASS, energy-gate pass rate, NEB convergence, and barrier plausibility across eight memory/RL configurations (full nine-row sweep, Table~\ref{si-tab:rl_ablation_full}).}
  \label{fig:agent45_rl_metrics}
\end{figure}

To prevent the per-step failures of this two-stage construction from recurring across runs, CatDT integrates a memory-augmented reinforcement architecture inspired by the Memento framework\cite{memento2025} that gives Agents~4 and~5 persistent external memory (Fig.~\ref{fig:agent45_rl_metrics}a).
The architecture is organised into four interacting modules.
The \emph{Network} module holds an episode-level $\varepsilon$-greedy bandit over workflow strategies and a parametric Q-network planner that, given the task and retrieved cases, predicts the next design action.
The \emph{Bank} module stores every completed trajectory as a structured case and periodically distills accumulated cases into reusable KnowledgeBank rules and SkillBank templates.
The \emph{Agent} module couples Agent~4 (endpoint proposer) with Agent~5 (deterministic gates plus reasoning-based plausibility checks) through a feedback-and-reject inner loop that iterates up to ten times before either advancing a candidate or declaring failure.
The \emph{Tool} module then runs the CHE/CP barrier tool on validated candidates, performs thermodynamic and kinetic validation, and emits the case-and-reward record that updates the Network parameters and refills the Bank.

Within a single iteration the four modules operate in sequence.
A structured query built from the reaction type, the transition signature, the pathway constraints, and the previous round's truncated feedback is first sent to the Bank, which returns the top-$k$ positive and negative cases together with the most relevant KnowledgeBank and SkillBank entries.
Retrieval starts with a unified non-parametric scorer and is replaced once the case bank exceeds 200 trajectories by a parametric retriever trained on supervised triples drawn from the bank itself, with the prompt format presented to the downstream agents unchanged (Supplementary Section~\ref{si-note:rl_loop}, Sec.~\ref{si-ssec:scorer}).
Agent~4 then consumes these retrieved cases, knowledge items, and skill templates on top of the tool-generated baseline steps and staging plan, while Agent~5 sees the same memory context augmented by a deterministic precheck summary and the UMA energy-gate report.
The Tool module's reward shaping ties the reinforcement signal to physical outcomes rather than syntactic validity (Methods), so that NEB convergence and barrier plausibility, not prompt compliance, drive the gradient, and the full iteration trace is then appended back to the case bank.

Within the Network module, the parametric retriever and the episode-level $\varepsilon$-greedy bandit form two decoupled reinforcement loops with orthogonal objectives, ``which past experience to surface'' and ``which workflow strategy to invoke'', with the bandit picking among conservative, balanced, and exploratory workflow strategies before each run and updating their Q-values from a composite episode reward (Supplementary Section~\ref{si-note:rl_loop}, Sec.~\ref{si-ssec:bandit}).
Because the case bank pools trajectories across all reaction types, both loops generalize naturally across chemistries: a CO$_2$-reduction case carrying a \texttt{*COOH}$\to$\texttt{*CO} transition can prime a CO-hydrogenation run, and negative cases from any reaction flag universal failure modes for the others.
The entire memory mechanism lives outside the LLM, with structured cases stored and retrieved rather than parameters updated, so upgrading the foundation model immediately upgrades reasoning quality without rebuilding the bank (Table~\ref{si-tab:rl_ablation_full}).

After training on 600 OC20 catalytic surfaces balanced across HER, OER, NRR, and CO$_2$RR (full protocol in Methods), the held-out PASS rate climbs monotonically from 41.2\% to 83.5\% over 60 episodes (Fig.~\ref{fig:agent45_rl_metrics}b). Two transient dips at episodes~25 and~45 coincide with bank-extraction events, after which PASS and reward resume their ascent, confirming that the reinforcement signal tracks chemically meaningful endpoint quality rather than prompt-level convenience.

The aggregate gain decomposes into the two independent learning loops (Fig.~\ref{fig:agent45_rl_metrics}c).
The parametric retriever raises its validation accuracy from a cold-start baseline to 0.92 as training pairs grow to ${\sim}$2{,}000, while the $\varepsilon$-greedy bandit progressively separates the Q-values of the three strategies, confirming that conservative and balanced edits outperform exploratory revisions.
Memory distillation counteracts retrieval dilution. The case bank grows approximately linearly to ${\sim}$600 entries by episode~60 (Fig.~\ref{si-fig:si_case_bank}), while the KnowledgeBank and SkillBank advance in the staircase pattern of the five-episode extraction cadence, reaching 171 knowledge items and 141 skill items respectively, and the reused-case fraction rises from 0.08 to 0.39 (Fig.~\ref{fig:agent45_rl_metrics}d), a quantitative signature of progressive compression of raw experience into reaction-agnostic abstractions.
Mean top-$k$ retrieval scores for knowledge and skill items sharpen in discrete steps phase-locked to the extraction schedule and, by the final checkpoint, match or exceed those of raw positive cases (Fig.~\ref{fig:agent45_rl_metrics}e).

A staged ablation at the final checkpoint resolves the contribution of each memory layer across the eight memory/RL configurations in the figure (Fig.~\ref{fig:agent45_rl_metrics}f), tabulated identically in Table~\ref{si-tab:rl_ablation_full}.
The stateless agent reaches a PASS rate of 49\%, the distilled banks alone (knowledge or skill) lift PASS to 52--54\%, and the raw case bank alone reaches 68\%, identifying trajectory recall as the single largest contributor.
Pairing the case bank with either distilled bank pushes PASS to 72--73\% (Table~\ref{si-tab:rl_ablation_full}), with the largest gains concentrated in NEB convergence and barrier plausibility, consistent with the hypothesis that abstracted geometric heuristics improve endpoint quality beyond what raw case recall can achieve.
The full reinforcement stack, with banks coupled to the episode-level bandit and the parametric retriever, reaches PASS 84\%, energy-gate 80\%, NEB convergence 69\%, and reasonable-barrier 65\%, with no single component dominating, confirming that trajectory recall, memory distillation, and learned retrieval address complementary failure modes (Table~\ref{si-tab:ablation}, Fig.~\ref{si-fig:si_pos_neg_ablation}).
With every link of the autonomous chain now built and self-improving, the remaining question is whether the assembled pipeline reproduces experimentally measured kinetic observables on real catalytic systems.

\FloatBarrier

\subsection{Gas--Solid Interface: End-to-End Validation on Thermal Catalysts}

\begin{figure}[H]
  \centering
  \makebox[\textwidth][c]{\includegraphics[width=1.1\textwidth]{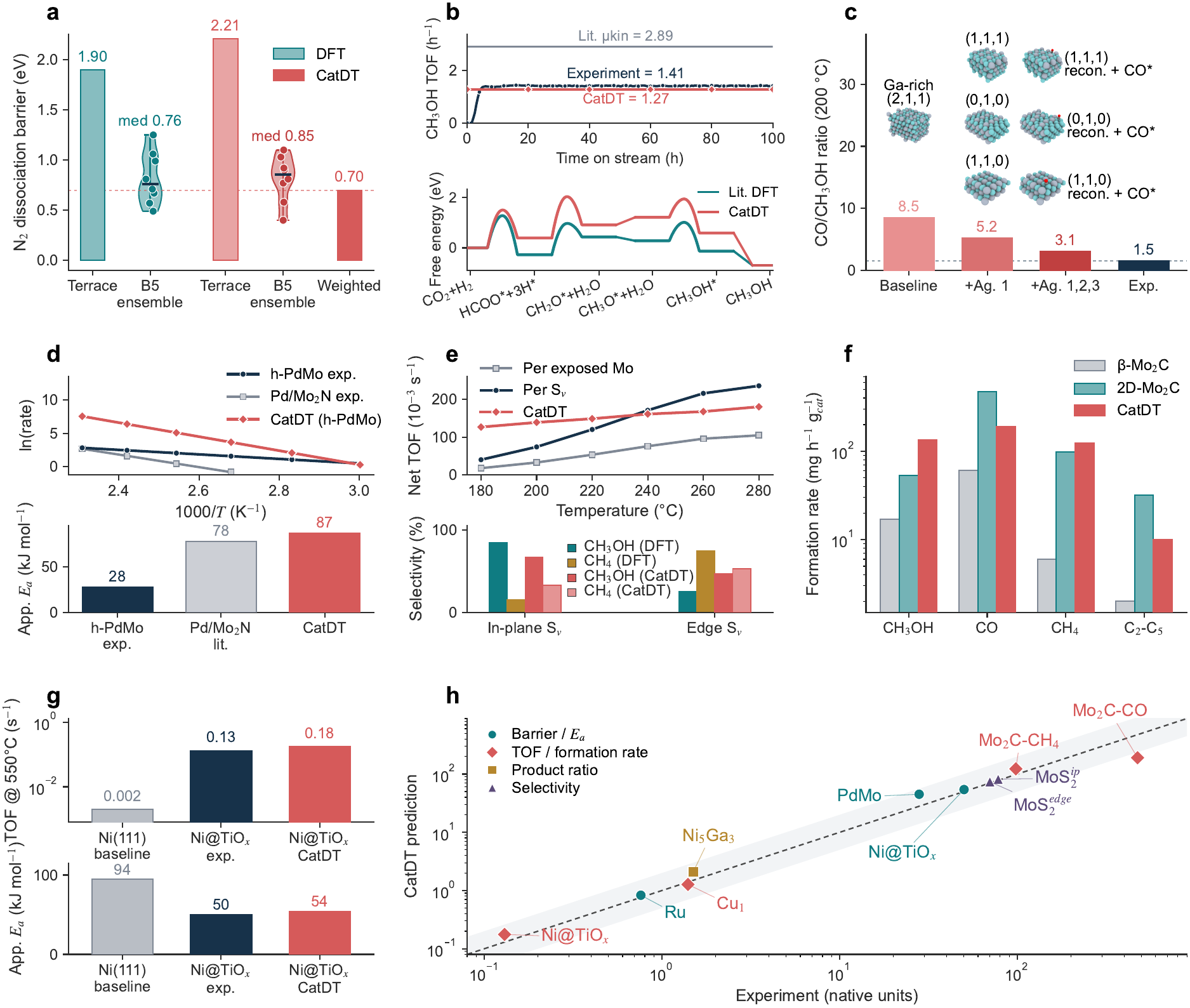}}
  \caption{\textbf{CatDT validation across seven gas--solid thermal catalysis benchmarks.}
  Colour key: teal/gold/olive/aubergine, DFT or literature microkinetics; black/grey, experiment; coral, CatDT.
  \textbf{a}.~Promoted Ru: N$_2$ dissociation barriers on the close-packed terrace and the B$_5$ step ensemble\cite{honkala2005ammonia}.
  \textbf{b}.~Cu$_1$-O$_3$/ZrO$_2$: methanol TOF (top) and CatDT vs literature free-energy diagram for the methanol branch (bottom)\cite{zhao2022cu1o3}.
  \textbf{c}.~Ni$_5$Ga$_3$: agent-ablation CO/CH$_3$OH ratio at 200~$^\circ$C\cite{studt2014niga}. CatDT surface structures rendered above each ablation stage by the CatDT visualisation module; the +Ag.~1 and +Ag.~1,2,3 columns each show the top-three Wulff-predicted facets, with the CO/CH$_3$OH ratio averaged over those three facets and over the AdsorbML-sampled binding sites per facet.
  \textbf{d}.~hcp-PdMo: Arrhenius plot and apparent $E_a$ against the Pd/Mo$_2$N control\cite{sugiyama2023pdmo}.
  \textbf{e}.~Vacancy-rich MoS$_2$: topology-weighted TOF vs.\ temperature (top) and in-plane-vs-edge selectivity reversal (bottom)\cite{hu2021svmos2}.
  \textbf{f}.~2D-Mo$_2$C: branch-resolved formation rates\cite{zhou2021mo2c}.
  \textbf{g}.~SMSI Ni@TiO$_x$/Al$_2$O$_3$: propylene TOF at 550~$^\circ$C (top) and apparent $E_a$ of the rate-determining C$_3$H$_7^*\!\to\!$C$_3$H$_6^*$ step (bottom)\cite{science2024nitiox}.
  \textbf{h}.~Parity plot of every benchmark observable; marker shape by observable type (circles, barriers/$E_a$; diamonds, TOF/formation rates; squares, product ratios; triangles, selectivities). The Ni$_5$Ga$_3$ CO/CH$_3$OH temperature sweep (168--249~$^\circ$C) is reported in Fig.~\ref{si-fig:si_ni5ga3_tdep}.}
  \label{fig:thermal_benchmark_suite}
\end{figure}

Geometrically valid NEB endpoints are necessary but not sufficient for a useful digital twin. The decisive test is whether the full CatDT stack (surface construction, pathway enumeration, barrier calculation, and microkinetic closure) reproduces experimentally measured kinetic observables on real catalytic systems with no manual intervention.
Because the pipeline is a serial chain (which facets are exposed, how the surface reconstructs, which pathways open, what the barriers are), an error at any link propagates downstream, so CatDT assigns each agent to guard exactly one link.
To stress-test this design we selected seven gas--solid thermal catalysis benchmarks of increasing complexity, each adding one challenge the previous system did not require (Fig.~\ref{fig:thermal_benchmark_suite}). In every case, CatDT received only a bulk crystal structure and a one-sentence natural-language reaction description, with all subsequent modeling performed autonomously.

The first benchmark probes the underlying tool chain on a known active site.
Promoted Ru catalysts owe their activity not to the dominant close-packed terrace, whose N$_2$ dissociation barrier is prohibitively high at 1.90~eV, but to the minority B5 step ensemble, where the barrier drops to 0.49--1.25~eV depending on the local coordination\cite{honkala2005ammonia}.
CatDT computes both the terrace and B5 barriers with UMA (Fig.~\ref{fig:thermal_benchmark_suite}a): the CatDT terrace barrier of 2.21~eV reproduces the DFT trend, and the CatDT B5 distribution with a median of 0.85~eV tracks the DFT ensemble at 0.76~eV with a mean absolute error of ${\sim}$0.1~eV across the eight local environments, a residual attributable to the UMA--DFT methodological difference.
The coverage-weighted effective barrier of 0.70~eV falls within the experimentally established B5 range, and propagating it through the coverage-dependent microkinetic model of Honkala et~al.\cite{honkala2005ammonia} reproduces the temperature dependence of the DFT-RPBE reference across the entire 320--440~$^\circ$C synthesis window, with an apparent activation energy of 0.94~eV against 1.06~eV from the original DFT-RPBE barriers (Fig.~\ref{si-fig:si_ru_b5_nh3}; Supplementary Section~\ref{si-ssec:ru_b5_nh3}).
The agreement of both the barrier ensemble and the propagated turnover frequencies with DFT references shows that the tool chain is quantitatively reliable before any higher-level agent capability is invoked.

Building on this foundation, the second benchmark asks whether CatDT can discover the correct reaction mechanism without human guidance.
On the Cu$_1$-O$_3$/ZrO$_2$ single-atom catalyst for CO$_2$ hydrogenation to methanol\cite{zhao2022cu1o3}, the active site is unambiguous, having been EXAFS-confirmed as quasiplanar Cu$_1$-O$_3$, but two competing pathways, HCOO*-mediated (formate) and COOH*-mediated (carboxyl), lead to different products.
The original study hand-selected both branches, whereas, starting from only the reaction string ``CO$_2$ hydrogenation to methanol'', UniMech autonomously enumerates candidates through its eight bidirectional bond-operation categories and a beam search with UMA energy ranking, and independently identifies the HCOO* route as energetically preferred.
The resulting free-energy profiles reproduce the literature DFT shape on both branches, with a systematic upward shift consistent with the UMA--PBE-D3 methodological difference.
Propagating the UniMech-discovered barriers through a Kozuch-style energetic-span analysis on the 453~K Gibbs profile of the original study yields a methanol TOF of 1.27~h$^{-1}$ (Fig.~\ref{fig:thermal_benchmark_suite}b, top), within 9\% of the experimental plateau of ${\sim}$1.4~h$^{-1}$ and a factor of two closer to experiment than the original microkinetic model, which overshot at 2.89~h$^{-1}$.  The milestone CatDT free-energy diagram underneath (Fig.~\ref{fig:thermal_benchmark_suite}b, bottom) overlays the autonomously discovered methanol-branch profile on the literature DFT path, making explicit which UniMech-discovered intermediates drive the TDI--TDTS shift behind the corrected TOF.
The improvement stems from an autonomously discovered shift in the TDI--TDTS pair that aligns the apparent activation energy with the experimental Arrhenius fit (full energetic-span breakdown in Supplementary Section~\ref{si-ssec:cu1o3_span}).

Facet prediction and surface reconstruction now enter the mechanism-discovery test, adding two upstream agents at once.
The Ni$_5$Ga$_3$ intermetallic\cite{studt2014niga} owes its low CO/CH$_3$OH ratio, distinct from that of the Cu/ZnO/Al$_2$O$_3$ industrial benchmark, to a site-partition effect: Ga-rich steps favor methanol, whereas Ni-rich steps produce CO and self-poison via strong CO binding at $\Delta E = -1.60$~eV.
The original study collapsed this physics onto a single fcc(211) surface and a one-descriptor volcano, whereas CatDT predicts the exposed facet distribution from the bulk (Agent~1, SurFF), models the 500~K surface reconstruction and adsorbate-induced CO poisoning (Agent~2, VSSR-MC, coupled to Agent~3 AdsorbDiff placement), and enumerates the full mechanism on both step types (Agent~M1).
A staged ablation at 200~$^\circ$C quantifies each agent's contribution (Fig.~\ref{fig:thermal_benchmark_suite}c). A baseline calculation on a static (211) facet yields a CO/CH$_3$OH ratio of 8.5, near the Cu/ZnO regime and far above the Ni$_5$Ga$_3$ experimental value of ${\sim}$1.5; the three CatDT surface states that produce this trend are rendered above the bars in Fig.~\ref{fig:thermal_benchmark_suite}c by the CatDT visualisation module (Ga-rich (211) baseline, Wulff-weighted Ni-rich exposure, and the CO-poisoned reconstructed surface). Adding Agent~1's Wulff multi-facet weighting reduces it to 5.2 by correcting the Ga-rich to Ni-rich exposure fraction, and further adding Agents~2 and~3 (thermal reconstruction plus CO-poisoning steady state) reduces it to 3.1 by deactivating Ni-rich sites, placing the prediction within a factor of two of experiment.
Each agent contributes an essential increment, and no single agent closes the gap.
The temperature-dependent CO/CH$_3$OH profile of the full pipeline remains well below the Cu/ZnO/Al$_2$O$_3$ benchmark across the entire 180--240~$^\circ$C window and tracks the experimental Ni$_5$Ga$_3$ trace (Fig.~\ref{si-fig:si_ni5ga3_tdep}), confirming that CatDT recovers the defining selectivity advantage of the intermetallic.

The three preceding systems each came with a literature DFT mechanism to compare against. The fourth benchmark removes that anchor entirely and asks whether CatDT can perform a genuine blind prediction on the hcp-PdMo intermetallic\cite{sugiyama2023pdmo}, for which the original study is purely experimental and reports only DRIFT-identified intermediates and an apparent activation energy of 28~kJ~mol$^{-1}$, roughly one-third of the 78~kJ~mol$^{-1}$ measured for the Pd/Mo$_2$N control, without any computed mechanism, barriers, or turnover frequencies.
From bulk Pd and Mo inputs alone, CatDT constructs the surface and UniMech discovers a reverse water-gas-shift plus CO hydrogenation pathway, CO$_2 \to$ CO* $\to$ HCO* $\to$ CH$_3$O* $\to$ CH$_3$OH, rather than the formate route favored by most methanol-synthesis catalysts, a non-trivial prediction that is independently corroborated by the DRIFT spectral assignments.
The Arrhenius fit to the CatMAP-derived rates yields an apparent $E_a$ of 87~kJ~mol$^{-1}$ (Fig.~\ref{fig:thermal_benchmark_suite}d), comparable in magnitude to the Pd/Mo$_2$N experimental value of 78~kJ~mol$^{-1}$ and systematically higher than the hcp-PdMo experimental value of 28~kJ~mol$^{-1}$, a residual offset consistent with the UMA--DFT systematic gap already observed for Ru and Cu$_1$-O$_3$.
This is the first benchmark in the suite where CatDT moves from reproducing known results to genuine blind prediction: it recovers the sub-100~kJ~mol$^{-1}$ regime of methanol synthesis over Mo-based intermetallics entirely from bulk inputs, with no prior DFT mechanism to anchor the prediction (Supplementary Section~\ref{si-ssec:pdmo_multi_p}; Fig.~\ref{si-fig:si_pdmo_multi_p}).

Having validated the pipeline on metallic catalysts, the fifth benchmark shifts to a fundamentally different material class.
Sulfur-vacancy-rich MoS$_2$\cite{hu2021svmos2} is a two-dimensional transition-metal dichalcogenide whose activity arises from defect sites rather than metal surfaces, and on which two defect topologies of the same material give opposite selectivities, with in-plane double sulfur vacancies favoring methanol and Mo-edge double vacancies favoring methane (barrier values in Supplementary Section~\ref{si-ssec:mos2_vacancy}).
CatDT builds both vacancy models, computes separate pathway--barrier sets via UniMech, and feeds them into a topology-weighted CatMAP calculation.
The weighted TOF falls between the two experimental normalizations, per exposed Mo and per sulfur vacancy, and reproduces the upward trend with temperature (Fig.~\ref{fig:thermal_benchmark_suite}e, upper), and the model also captures the selectivity reversal between in-plane and edge vacancies (Fig.~\ref{fig:thermal_benchmark_suite}e, lower), confirming that UMA and UniMech transfer to non-metallic two-dimensional materials.

The sixth system stresses the pipeline's multi-product capability.
Two-dimensional Mo$_2$C\cite{zhou2021mo2c} presents the broadest product spectrum in this suite, simultaneously producing CO, CH$_3$OH, CH$_4$, and C$_2$--C$_5$ hydrocarbons at 230~$^\circ$C and 25~bar over the reduced Mo-terminated surface, and exercises every CatDT capability. Agent~1 picks the Mo-terminated (0002) facet, consistent with the SAED data of the original study. Agents~2 and~3 model the surface after reductive removal of T$_x$ termination groups, and UniMech then enumerates all product branches, including the C--C coupling steps required for C$_2$--C$_5$ formation that lie beyond the scope of narrower C/H/O/N frameworks.
CatDT correctly identifies CO as the dominant product and C$_2$--C$_5$ as the least abundant, with branch-resolved formation rates within the experimental range (Fig.~\ref{fig:thermal_benchmark_suite}f; Table~\ref{si-tab:mo2c_rates}; Supplementary Section~\ref{si-ssec:mo2c_table}).

The seventh and most demanding benchmark is SMSI-encapsulated Ni@TiO$_x$/Al$_2$O$_3$ for propane dehydrogenation (PDH)\cite{science2024nitiox}.
It tests whether the pipeline can capture an interfacial site whose activity emerges from TiO$_x$ migration over Ni rather than from a bulk-defined facet.
Starting from bulk Ni and TiO$_2$, Agents~1--3 autonomously reproduce the VSSR-MC evidence for Ti$^{3+}$ overlayer formation on Ni(111).
The UniMech--NEB chain on the reconstructed surface then delivers the two principal kinetic observables simultaneously (Fig.~\ref{fig:thermal_benchmark_suite}g).
The predicted propylene TOF at 550~$^\circ$C is 0.16~s$^{-1}$, matching the experimental 0.13~s$^{-1}$ within 25\%\cite{science2024nitiox}.
It also reproduces the ${\sim}$65-fold enhancement over the Ni(111) baseline from the SMSI-reconstructed interface (Fig.~\ref{fig:thermal_benchmark_suite}g, top).
The apparent activation energy for the rate-determining C$_3$H$_7^*\!\to\!$C$_3$H$_6^*$ step is 54~kJ~mol$^{-1}$ (50~kJ~mol$^{-1}$ experimentally).
This recovers $\sim$90\% of the SMSI-induced reduction from the 94~kJ~mol$^{-1}$ Ni(111) baseline (Fig.~\ref{fig:thermal_benchmark_suite}g, bottom; Supplementary Section~\ref{si-ssec:nitiox_ea}; Fig.~\ref{si-fig:si_ni_tiox_fe}).
These two observables together show that CatDT generalizes from oxide-supported single atoms, as in Cu$_1$-O$_3$ above, to metal-supported SMSI overlayers without any system-specific tuning.
The one observable for which CatDT still falls short is propylene selectivity: the predicted C$_3$H$_6$/C$_2$H$_y$ ratio underestimates experiment because the multi-path KMC is not yet coupled to UniMech's Sabatier gating, an integration left to follow-on work (Supplementary Section~\ref{si-ssec:nitiox_selectivity}).

Aggregated across all seven benchmarks, the ten directly comparable CatDT observables sit inside a $\sim$2$\times$ band around the 1:1 experiment line, spanning four decades of dynamic range and uniformly distributed across barriers, turnover frequencies and formation rates, product ratios, and selectivities (Fig.~\ref{fig:thermal_benchmark_suite}h; Supplementary Section~\ref{si-note:benchmark_details}, Secs.~\ref{si-ssec:ru_b5_nh3}--\ref{si-ssec:benchmark_summary}; Figs.~\ref{si-fig:si_ru_b5_nh3}--\ref{si-fig:si_ni_tiox_fe}; Tables~\ref{si-tab:mo2c_rates}--\ref{si-tab:benchmark_summary}).

Taken together, these seven systems, ordered by increasing demand on the pipeline, establish that a single autonomous pipeline reproduces qualitative trends and, in most cases, quantitative magnitudes of experimentally measured kinetic observables across seven distinct material classes without system-specific parameter tuning.
The Ni$_5$Ga$_3$ ablation confirms that each upstream agent contributes an essential increment to the final observable, and the hcp-PdMo benchmark shows that CatDT can predict kinetic properties for systems lacking any prior computational mechanism.
Where quantitative discrepancies remain, they point to specific, addressable modeling simplifications rather than fundamental failures of the architecture (Supplementary Section~\ref{si-note:benchmark_details}; Table~\ref{si-tab:benchmark_summary}).
Recovering these seven benchmarks within a factor of two is the prerequisite for the more ambitious test of deploying the same infrastructure as a closed-loop discovery engine, where the target shifts from matching known kinetics to surfacing candidates that improve on the industrial state of the art.

\FloatBarrier

\subsection{Cross-System Intelligent Design of Non-Precious PDH Catalysts}

\begin{figure}[H]
  \centering
  \makebox[\textwidth][c]{\includegraphics[width=1.1\textwidth]{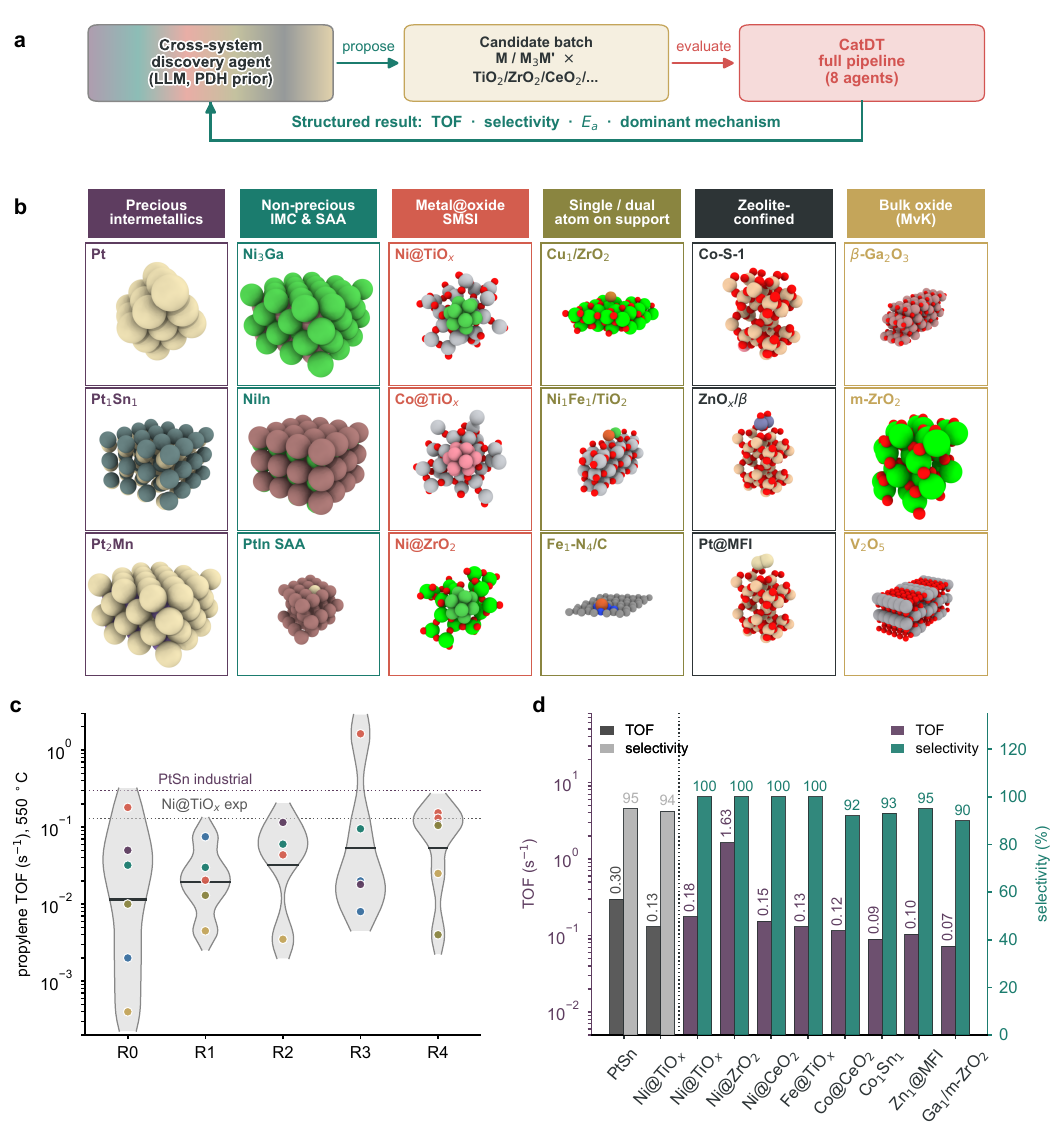}}
  \caption{\textbf{Cross-system intelligent design of non-precious PDH catalysts with CatDT.}
  \textbf{a}.~Closed-loop discovery system coupling the Discovery Agent, the candidate batch, and the CatDT pipeline.
  \textbf{b}.~PDH design portfolio: six catalyst families (precious intermetallics; non-precious intermetallics and single-atom alloys; metal@oxide SMSI overlayers; single- or dual-atom sites on oxide or carbon supports; zeolite-confined sites; bulk reducible oxides) each represented by three characteristic structures rendered with the CatDT visualisation module.
  \textbf{c}.~Per-round propylene TOF violin across the five Discovery-Agent rounds (R0--R4); each candidate's point colour matches its panel-b family. Dashed horizontal lines mark the Ni@TiO$_x$ experimental and PtSn industrial references.
  \textbf{d}.~TOF (left purple log axis) and C$_3$H$_6$ selectivity (right green linear axis) across experimental benchmarks (dark/light gray bars, left of the divider) and CatDT predictions (purple/green bars, right of the divider).
  The corresponding Ni@TiO$_x$ PDH free-energy profile is provided in Fig.~\ref{si-fig:si_ni_tiox_fe}.}
  \label{fig:pdh_discovery}
\end{figure}

Having validated CatDT across seven gas--solid benchmarks, we deploy it as a closed-loop catalyst-design engine for the cross-system discovery of non-precious propane-dehydrogenation (PDH) catalysts.
A \emph{Discovery Agent} seeded with PDH domain priors proposes a batch of candidate materials, each of which is evaluated by the full CatDT pipeline, and the structured result, comprising turnover frequency (TOF), propylene selectivity, apparent activation energy, and dominant mechanism, is returned as feedback for the next round of selection (Fig.~\ref{fig:pdh_discovery}a).

The design space surveyed across the PDH literature spans six structurally distinct catalyst families (Fig.~\ref{fig:pdh_discovery}b): precious intermetallics (Pt$_1$Sn$_1$\cite{motagamwala2021pt1sn1}, PtGa\cite{sun2021ptga}, and Pt$_2$Mn\cite{rochlitz2022pt2mn}), non-precious intermetallics and single-atom alloys (Ni$_3$Ga\cite{siddiqi2017ni3ga,nakaya2018ni3ga}, NiIn\cite{furukawa2018niin}, Ni$_3$V\cite{han2024ni3v_dft} and PtIn SAA\cite{zhou2025ptin}), SMSI metal@oxide overlayers (Ni@TiO$_x$\cite{science2024nitiox}, with Co@TiO$_x$ and Ni@ZrO$_2$ agent-proposed), single- or dual-atom sites on oxide or carbon supports (Cu$_1$/ZrO$_2$\cite{zhao2022cu1o3}, Ni$_1$Fe$_1$/TiO$_2$\cite{yao2025ni1fe1} and Ru$_1$/NC\cite{fe_n4_c_2022}), zeolite-confined active sites (CoS-1\cite{zhou2025cos1}, ZnO$_x$/$\beta$\cite{zhao2021znox}, Pt@MFI\cite{park2025ptmfi}, and CoO$_x$/S-1\cite{liu2025coox_s1}), and bulk reducible oxides operating via the Mars--van Krevelen mechanism ($\beta$-Ga$_2$O$_3$\cite{ga2o3_acscat2021}, m-ZrO$_2$\cite{zhang2018zro2}, V$_2$O$_5$\cite{vox_zro2_2021}, and the tandem In$_2$O$_3$--Pt system\cite{yan2021in2o3pt}).
Each family covers a qualitatively different active-site motif and is represented by three characteristic structures.
The anchor for the present study is Ni@TiO$_x$/Al$_2$O$_3$\cite{science2024nitiox}, the SMSI-encapsulated benchmark already validated in Fig.~\ref{fig:thermal_benchmark_suite}g.

Round~0 probes a cross-family seed batch (Fig.~\ref{fig:pdh_discovery}c, R0): starting from the Ni@TiO$_x$ anchor and pairing it with one representative each from the five remaining families of Fig.~\ref{fig:pdh_discovery}b, CatDT returns six propylene TOFs spanning more than two orders of magnitude.
The Ni@TiO$_x$\cite{science2024nitiox} prediction sits at the upper edge of this distribution and approaches the PtSn industrial window, while the other five seed candidates trail the benchmark by up to three orders of magnitude, indicating to the Discovery Agent that no single family is sufficient on its own.
After each round, the agent ingests the structured CatDT report (rate-determining barriers, dominant mechanism, H coverage, and selectivity decomposition) and writes a short reflection diagnosing why specific candidates over- or underperformed. It then turns that diagnosis into the next batch: high-performing motifs (SMSI overlayer, non-precious IMC, single-atom site on a reducible oxide) are amplified, low-performing motifs (zeolite-confined Co or Zn at high temperature, bulk MvK oxides for which the rate-determining C--H step lies far above the lattice-O activation step) are pruned, and new analogues recombine surviving motifs across families.

This reflection loop drives the trend visible in Fig.~\ref{fig:pdh_discovery}c.
From R0 through R2, the per-round geometric-mean TOF rises monotonically as the agent converges on the SMSI overlayer family and its non-precious-IMC siblings, drawing successive 3$d$-metal/TiO$_x$ analogues into each new batch.
The decisive step occurs in R3: pushed by the reflection that ``the TiO$_x$ overlayer suppresses deep dehydrogenation, but the active interface charge transfer is tunable through the support oxide,'' CatDT proposes Ni@ZrO$_2$ and recovers a CatMAP--KMC propylene TOF of $1.63~\text{s}^{-1}$ with $\sim$100\% selectivity, about five times the PtSn industrial benchmark and the strongest non-precious PDH candidate of the loop.
R4 then extends the sweep to other reducible-oxide overlayers (Ni@CeO$_2$, Fe@TiO$_x$) and reference families (Pt, Ni$_1$Fe$_1$/TiO$_2$), several of which cluster just below the PtSn industrial line at $\sim$100\% selectivity, after which the agent's reflection signals convergence and terminates the loop.
A parallel sweep of more strongly reducible Group-V/VI Ni@oxide overlayers fails the Stage-A barrier gate or returns below-threshold TOFs and is therefore omitted from the Fig.~\ref{fig:pdh_discovery}c violin (full per-round inventory with the specific candidate TOFs and the failed-overlay diagnosis in Supplementary Section~\ref{si-note:pdh_rounds_detail}).

For the Ni@TiO$_x$ anchor, the full CatDT pipeline delivers a two-step PDH free-energy profile (Fig.~\ref{si-fig:si_ni_tiox_fe}) whose rate-determining C$_3$H$_7^*\!\to\!$C$_3$H$_6^*$ barrier matches the Science~2024 apparent activation energy to within the UMA universal-potential uncertainty\cite{uma2025,science2024nitiox} and lies well below the clean Ni(111) baseline.
The downstream CatMAP\cite{medford2015catmap} closure places the prediction inside the PtSn industrial TOF window (Fig.~\ref{fig:pdh_discovery}d) and recovers the Ni@TiO$_x$ experimental TOF and propylene selectivity of Kim~et~al.\cite{science2024nitiox}, with the Ni@ZrO$_2$ winner sitting roughly five times above the PtSn reference and the next-best SMSI overlayer winners (Ni@CeO$_2$ at 0.15~s$^{-1}$, Fe@TiO$_x$ at 0.13~s$^{-1}$) clustering just below the PtSn line, all at $\sim$100\% predicted propylene selectivity.
This pattern indicates that the SMSI overlayer suppresses the deep-dehydrogenation branch through a site-averaged H coverage that transfers from the Ni@TiO$_x$ anchor to the agent-proposed Ni@ZrO$_2$, Ni@CeO$_2$ and Fe@TiO$_x$ analogues.

A non-precious replacement for the PtSn industrial benchmark must also be materially cheaper, and the SMSI-overlayer motif favoured by the discovery loop is precisely the architecture that minimises precious-metal content by design.
Under representative industrial loadings and May~2026 commodity prices (full price table and loading assumptions in Supplementary Section~\ref{si-note:pdh_cost}), the leading agent-proposed analogues fall 20--130$\times$ below the PtSn benchmark in raw material cost, and the cost-per-activity figure of merit $F=c_\text{cat}/\mathrm{TOF}$ (Fig.~\ref{si-fig:si_pdh_cost}b) places Ni@ZrO$_2$ about $162\times$ below the PtSn baseline, with Fe@TiO$_x$ and the CatDT Ni@TiO$_x$ prediction both $\sim$56$\times$ below.

These results (Fig.~\ref{fig:pdh_discovery}c--d and the cost analysis in Fig.~\ref{si-fig:si_pdh_cost}) show CatDT operating as a closed-loop, cross-family PDH design engine: reflection-guided batch selection spans all six families, converges on Ni@ZrO$_2$ as the leading non-precious candidate at roughly thirty times lower raw-material cost than the PtSn industrial reference\cite{motagamwala2021pt1sn1}, and keeps Ni@TiO$_x$\cite{science2024nitiox} as a quantitative SMSI-PDH benchmark.
On a single RTX~4090 (48~GB) GPU, the per-structure wall time spans roughly 0.5--6~h and stratifies by structural complexity: single- and dual-atom sites on oxide supports clear the full bulk-to-KMC pipeline in well under 1.5~h, bulk reducible oxides in 1.4--2.2~h, the intermetallic and metal\,@\,oxide SMSI candidates (whose VSSR-MC reconstruction sweep dominates the cost) in 2.4--4.6~h, and the zeolite-confined candidates, whose large MFI/CHA/$\beta$ supercells are the most expensive per evaluation, in 4.7--5.7~h (Fig.~\ref{si-fig:si_throughput}; full per-family discussion in Supplementary Section~\ref{si-note:throughput}). The entire six-family, twenty-candidate-plus discovery loop therefore fits inside a single-GPU-week with no expert intervention.
A detailed kinetic dissection of the leading non-precious candidates emerging from batches R1--R4, together with their synthesis-feasibility assessment, lies outside the scope of the present infrastructure paper and is the subject of dedicated follow-on work.
The discovery loop therefore exercises every layer of the CatDT stack end-to-end, returning a ranked, cost-aware candidate pool from no input other than the reaction string.

\FloatBarrier

\section{Conclusion}

CatDT demonstrates that the full theoretical catalysis pipeline, from a bulk crystal to experimentally comparable kinetic observables, can be made autonomous, self-improving, and quantitatively predictive.
Two innovations close the structural barriers that have constrained the field: UniMech finds energy-ranked pathways with 97 evaluations where exhaustive enumeration requires 455{,}000 ($>$10$^3\times$ reduction on C$_6$ Fischer--Tropsch), and a memory-augmented reinforcement loop lifts NEB endpoint quality from 41\% to 84\% PASS across 600 diverse surfaces, distilling raw experience into 171 reusable knowledge items and 141 skill templates.
Across seven gas--solid benchmarks the autonomous pipeline reproduces every directly comparable kinetic observable to within a factor of two of experiment over four decades of dynamic range, and as a closed-loop design engine it returns non-precious propane-dehydrogenation candidates that rival the Pt-based industrial benchmark, with the agent-proposed Ni@ZrO$_2$ SMSI overlayer reaching a CatMAP--KMC propylene TOF of $1.63~\text{s}^{-1}$ at $\sim$100\% selectivity.

Because the agent--tool separation decouples reasoning from computation, CatDT functions as upgradable infrastructure: replacing UMA with a more accurate potential, upgrading the foundation model, or registering tool backends for photocatalysis or solid--solid interfaces requires no architectural change, and the accumulated case, knowledge, and skill banks persist across every such upgrade.
More broadly, scientific reliability in agent systems depends on the engineered verification, memory, and tool-orchestration surface around the model rather than on the model itself, so correctness becomes a property of the system rather than of any single invocation, and each run measurably improves the twin's predictive accuracy.
Coupled to an autonomous wet laboratory, in which robotic synthesis, in-situ characterization, and kinetic testing close the loop on the physical catalyst\cite{boiko2023coscientist}, the same harness can serve as the in-silico half of a fully self-driving catalyst-discovery cycle, where agent-proposed candidates are computationally screened, experimentally validated, and fed back into the digital twin within hours rather than months.
This convergence of autonomous theory and autonomous experiment, anchored on a harness that grows with every run, points to a near term in which designing a new heterogeneous catalyst becomes a continuous feedback loop between machine and reactor rather than a multi-year human campaign.

\section{Methods}

\subsection{Universal machine-learning potentials and thermodynamic corrections}\label{subsec:ml_potentials}

Gas--solid calculations use \textbf{UMA (uma-s-1p1)}\cite{uma2025}, trained on OC20/22/25\cite{chanussot2021oc20,tran2022oc22,shuaibi2025oc25}. The same potential is used for SurFF-backed facet ranking, AdsorbDiff placement scoring, VSSR-MC reconstruction energies, UniMech cache evaluations, the Agent~5 pre-NEB energy gate, and \textbf{two-phase CI-NEB} barrier searches. CI-NEB first relaxes an initial minimum-energy path with FIRE and then enables a climbing image; barriers are extracted as $E_\text{a}=E_\text{TS}-E_\text{reactant}$. Non-convergent or non-physical paths are retried through the adaptive NEB ladder and finally screened by the post-NEB validator (Table~\ref{si-tab:neb_attempt_ladder}, Fig.~\ref{si-fig:neb_attempt_ladder}, Table~\ref{si-tab:neb_validator}; Supplementary Section~\ref{si-note:endpoints}).
Liquid--solid calculations require a \textbf{universal constant-potential machine-learning potential}: the model must respond to changing electron count and electrode potential while retaining the chemical transferability expected from a foundation potential. CP-MACE\cite{wang2025cpmace} provides the prototype constant-potential MACE architecture, but is not a universal off-the-shelf backend for every electrochemical interface; each target system still requires separate training. To make this practical, we developed a CP-MLIP active-learning system that \textbf{modifies some influential universal-potential architectures and GNNs} (MACE, UMA, EquiformerV2/ESEN, MatterSim, NEP, SchNet, DimeNet++, and PaiNN) to be CP-aware through architecture-specific electron conditioning and potential-prediction heads. The system starts from OC25-derived subsets and trains CP-aware models through iterative active learning. We will release all model parameters trained with our framework on Hugging Face.

\subsection{Thermodynamic corrections and molecule database}

Every electronic energy is converted to a \textbf{Gibbs free energy} using $\Delta G(T)=\Delta E+\Delta\mathrm{ZPE}-T\Delta S$. To keep this conversion consistent across mechanism search, endpoint construction, and kinetic closure, CatDT uses an internal \textbf{569-species molecule database} containing 464 surface adsorbates and 105 gas-phase references relevant to the benchmark reactions (Fig.~\ref{si-fig:thermo_db_stats}, Table~\ref{si-tab:thermo_db_composition}). The adsorbate pool is initialized from the Fairchem adsorbate set and expanded with reaction-specific gas-phase reactants, products, and reference molecules needed for the thermal benchmarks.
Each database entry is relaxed with UMA in a 30~\AA{} cubic cell and characterized with \textbf{ASE-Vibrations} at the same level of theory. Gas-phase species are evaluated with ideal-gas thermochemistry at the operating $(T,P)$, whereas adsorbed species use harmonic vibrational corrections. UniMech reads the same database when assigning cached intermediate free energies, Agent~4/5 use it to build chemically consistent product geometries, and Agent~6 uses it to pass temperature-corrected barriers and intermediate free energies into \textbf{CatMAP}. Database construction and numerical defaults are provided in Supplementary Sections~\ref{si-note:thermo_db} and~\ref{si-note:hyperparams} (Fig.~\ref{si-fig:thermo_db_stats}; Tables~\ref{si-tab:thermo_db_composition} and~\ref{si-tab:hyperparams}).

\subsection{UniMech tree search: shared cache, beam pruning, and MCTS}

UniMech runs three explicit stages. First, an LLM-seeded generator proposes literature-like mechanisms while a \textbf{deterministic RDKit enumerator} expands the reaction graph over sixteen directional, element-agnostic bond operations (Table~\ref{si-tab:unimech_ops}). Second, RDKit valence checks remove chemically invalid intermediates and an LLM plausibility pass retains the top-ranked candidates. Third, a \textbf{shared UMA energy cache} evaluates only frontier intermediates and ranks complete routes. The default search backend is \textbf{best-first beam search} (width~8, depth~8) with sibling pruning at $\Delta_\text{prune}=0.30$~eV; when a route must pass through transiently uphill intermediates, the backend switches to \textbf{MCTS} with $c_\text{UCB}=1.4$, 100 iterations, and a 40-evaluation budget. Full pseudocode, operator definitions, and all search defaults are in Supplementary Section~\ref{si-note:unimech} (Fig.~\ref{si-fig:si_unimech_detail}; Tables~\ref{si-tab:pathways_ch4}--\ref{si-tab:pathways_summary}; Figs.~\ref{si-fig:si_ch4_network}--\ref{si-fig:si_phase1_care}) and Table~\ref{si-tab:hyperparams}.

\subsection{Agent--tool harness and foundation-model assignment}

CatDT is implemented in CAMEL\cite{li2023camel} with strict agent--tool separation: deterministic \texttt{FunctionTool} calls produce every numerical result, while agents plan, validate, and route tasks. The deployed system contains 27 tools and eight agents (Supplementary Section~\ref{si-note:architecture}; Tables~\ref{si-tab:tools_full}--\ref{si-tab:agents}) orchestrated by Agent~7 through a workflow state machine with checkpoints and structured reports. The multi-agent system (MAS) is \textbf{foundation-model-agnostic}: every per-agent model choice is a configuration entry rather than a code dependency, so any agent backend can be swapped without changing the scientific tools or downstream parsing. By default, Agents~1, 2, 3, and~6 call \textbf{DeepSeek~V4~Pro} for high-throughput surface, adsorption, reconstruction, and kinetics control; Agents~4, 5, and~M1 call \textbf{GPT-5.5} for geometry design, validation auditing, and mechanism reasoning; and Agent~7 directly calls \textbf{Codex} (with \textbf{GPT-5.5} as its backend) for orchestration and structured output. Token budgets, watchdogs, retries, truncation caps, and strict JSON-schema gates are specified in Table~\ref{si-tab:harness_budgets} (Supplementary Section~\ref{si-note:harness}).

\subsection{Memento-style reinforcement training}

The Agent~4/5 memory loop follows \textbf{Memento-style external case retrieval}\cite{memento2025}. Each iteration stores a \textbf{structured trace} containing the reaction type, transition signature, endpoint-design JSON, Agent~5 diagnostics, energy-gate report, NEB outcome, and scalar reward. Retrieval begins with a non-parametric scorer over positive cases, negative cases, KnowledgeBank items, and SkillBank templates; after 200 trajectories, the scorer is replaced by a \textbf{parametric retriever} trained on supervised triples mined from the case bank. Training uses \textbf{600 OC20 catalytic surfaces} balanced across HER, OER, NRR, and CO$_2$RR; each episode aggregates ten full design--validation runs, with validation every five episodes on a fixed 50-case holdout. The iteration reward is $r_\text{iter}=0.4\rho_\text{conv}+0.6\rho_\text{barrier}$, clipped to zero when either NEB convergence or barrier plausibility vanishes. A separate $\varepsilon$-greedy bandit chooses among conservative, balanced, and exploratory workflow strategies, with episode-level credit assignment composed of validation success, NEB quality, and step-set completeness. Supplementary Section~\ref{si-note:rl_loop} expands two-stage endpoint construction, scorer design, rewards, bandit updates, protocol, and staged ablations (Secs.~\ref{si-ssec:two_stage}--\ref{si-ssec:rl_ablation}; Figs.~\ref{si-fig:si_case_bank}--\ref{si-fig:si_pos_neg_ablation}; Tables~\ref{si-tab:rl_ablation_full}--\ref{si-tab:ablation}); retriever, reward, bandit, and memory-bank schemas appear in Supplementary Section~\ref{si-note:memento_schema} and Table~\ref{si-tab:memento_schema}.

\FloatBarrier

\BackmatterHeading{Data Availability}

\indent The RL-training trajectories used in Section~\ref{subsec:memory_rl}, the 569-species thermodynamic-correction database described in Section~\ref{subsec:ml_potentials}, and the CP-aware pretrained model parameters referenced in Section~\ref{subsec:ml_potentials} will be made available on Hugging Face upon publication.

\BackmatterHeading{Code Availability}

\indent The CatDT package will be made available at \url{https://github.com/AI4QC/catdt} upon publication.

\BackmatterHeading{Acknowledgements}

\indent This research is done with the support from HKUST Start-up Funding with a GPU cloud computing infrastructure provided by ScitiX (Scitix (SGP) Tech Pte Ltd, Singapore). Z.Z thanks the support from Hong Kong Special Administrative Region Government Scholarship Fund Scholarship for Outstanding Performance. We also thank Mr. Hiuyu Ting from the Department of Chemistry, HKUST, for the helpful discussion.

\BackmatterHeading{Competing Interests}

\indent The authors declare no competing interests.

\bibliography{refs}

\end{document}